\documentclass[aps,pra,twocolumn,showpacs,superscriptaddress,floatfix]{revtex4-1}
\usepackage{graphicx,amsmath,amssymb}
\usepackage[usenames]{color}
\usepackage[dvipsnames]{xcolor}
\usepackage[unicode=true,pdfusetitle,
 bookmarks=false,
 breaklinks=false,pdfborder={0 0 1},backref=false,colorlinks,urlcolor=Black, linkcolor=blue,citecolor=blue]
 {hyperref}
\begin{document}

\title{Spin Winding and Topological Nature of Transitions in Jaynes-Cummings Model with Stark Non-linear Coupling
}
\author{Zu-Jian Ying }
\email{yingzj@lzu.edu.cn}
\affiliation{School of Physical Science and Technology, Lanzhou University, Lanzhou 730000, China}
\affiliation{Key Laboratory for Quantum Theory and Applications of MoE, Lanzhou Center for Theoretical Physics, and Key Laboratory of Theoretical Physics of Gansu Province, Lanzhou University, Lanzhou, Gansu 730000, China}

\begin{abstract}
Besides exploring novel transition patterns, acquiring a full understanding of the transition nature is an ultimate pursuit in studies of phase transitions. The fundamental models of light-matter interactions manifest single-qubit topological phase transitions,
which is calling for an analytical demonstration apart from numerical studies.
We present a rigorous study for topological transitions in Jaynes-Cummings Model generally with Stark non-linear Coupling. In terms of the properties of Hermite polynomials, we show that the topological structure of the eigen wave function has an exact correspondence to the spin winding by nodes, which yields a full spin winding without anti-winding nodes. The spurious fractional contribution to the winding number of the winding angle at infinity is found to be actually integer. Thus, the phase transitions in the model have a nature of topological phase transitions and the excitation number is endowed as a topological quantum number. The principal transition establishes a paradigmatic case that a transition is both symmetry-breaking Landau class of transition and symmetry-protected topological class of transition simultaneously, while conventionally these two classes of transitions are incompatible due to the contrary symmetry requirements. We also give an understanding for the origin of unconventional topological transitions in the presence of counter-rotating terms. Our results may provide a deeper insight for the few-body phase transitions in light-matter interactions.
\end{abstract}
\pacs{ }
\maketitle


\section{Introduction}

The recent years have witnessed both theoretical progesses~\cite{Braak2011,Solano2011,Boite2020,Liu2021AQT} and experimental adavances~\cite{Ciuti2005EarlyUSC,Aji2009EarlyUSC,Diaz2019RevModPhy,Kockum2019NRP,Wallraff2004,Gunter2009,
Niemczyk2010,Peropadre2010,FornDiaz2017,Forn-Diaz2010,Scalari2012,Xiang2013,Yoshihara2017NatPhys,Kockum2017,Bayer2017DeepStrong}
in the frontiers of
light-matter interactions. In this context, especially with the entrance
into the era of ultra-strong~\cite{
Ciuti2005EarlyUSC,Aji2009EarlyUSC,Diaz2019RevModPhy,Kockum2019NRP,Wallraff2004,Gunter2009,Niemczyk2010,
Peropadre2010,FornDiaz2017,Forn-Diaz2010,Scalari2012,Xiang2013,Yoshihara2017NatPhys,Kockum2017}
and deep-strong~\cite{Yoshihara2017NatPhys,Bayer2017DeepStrong} couplings, few-body
quantum phase transitions (QPTs) have become practically relevant and attracted a
special attention~\cite{
Liu2021AQT,Ashhab2013,Ying2015,Hwang2015PRL,Ying2020-nonlinear-bias,Ying-2021-AQT,LiuM2017PRL,Hwang2016PRL,Irish2017,
Ying-gapped-top,Ying-Stark-top,Ying-Spin-Winding,Ying-2018-arxiv,Ying-Spin-Winding,Grimaudo2022q2QPT}
among the massive efforts~\cite{
Braak2011,Solano2011,Boite2020,Liu2021AQT,Diaz2019RevModPhy,Kockum2019NRP,Rabi-Braak,Braak2019Symmetry,
Wolf2012,FelicettiPRL2020,Felicetti2018-mixed-TPP-SPP,Felicetti2015-TwoPhotonProcess,Simone2018,Alushi2023PRX,
Irish2014,Irish2017,Irish-class-quan-corresp,
PRX-Xie-Anistropy,Batchelor2015,XieQ-2017JPA,
Hwang2015PRL,Bera2014Polaron,Hwang2016PRL,Ying2015,LiuM2017PRL,Ying-2018-arxiv,Ying-2021-AQT,Ying-gapped-top,Ying-Stark-top,Ying-Spin-Winding,Grimaudo2022q2QPT,Grimaudo2023-Entropy,
CongLei2017,CongLei2019,Ying2020-nonlinear-bias,LiuGang2023,ChenQH2012,
e-collpase-Garbe-2017,e-collpase-Duan-2016,Garbe2020,Rico2020,
Garbe2021-Metrology,Ilias2022-Metrology,Ying2022-Metrology,
Boite2016-Photon-Blockade,Ridolfo2012-Photon-Blockade,Li2020conical,
Ma2020Nonlinear,
ZhangYY2016,ZhengHang2017,Yan2023-AQT,Zheng2017,Chen-2021-NC,Lu-2018-1,Gao2021,PengJie2019,Liu2015,Ashhab2013, ChenGang2012,FengMang2013,Eckle-2017JPA,Maciejewski-Stark,Xie2019-Stark,Casanova2018npj,HiddenSymMangazeev2021,HiddenSymLi2021,HiddenSymBustos2021,
JC-Larson2021,Stark-Cong2020,Cong2022Peter,Stark-Grimsmo2013,Stark-Grimsmo2014} in the digologue between mathematics and physics~\cite{Solano2011}
inspired by the milestone finding of integrability of the fundamental
light-matter-interaction model~\cite{Braak2011}. Few-body QPTs are fascinating not only
because of its exhibition of critical and universal behaviors~\cite%
{LiuM2017PRL,Hwang2015PRL,Hwang2016PRL,Ying-2021-AQT,Ying-Stark-top,Irish2017} as in
many-body systems~\cite{LiuM2017PRL,Irish2017} but also due to its high controbility and tunability which
show advantages in applications such as in quantum metrology~\cite%
{Garbe2020,Garbe2021-Metrology,Ilias2022-Metrology,Ying2022-Metrology}.

Phase transition (PT) is a ubiquitous phenomenon in our physical world. The
investigation of PTs is a field full of challenges, whereas surprising
discoveries may also be often encountered. Exploring novel patterns of PTs
and seeking a full understanding of PTs have always been a goal. In this
regard, the well-known Landau theory~\cite{Landau1937} made a breakthrough in understanding
traditional phase transitions by realizing that a PT is
associated with some symmetry breaking, while another essentially different
class of PT is the topological phase transition (TPT)~\cite{Topo-KT-transition,Topo-KT-NoSymBreak,Topo-Haldane-1,Topo-Haldane-2,Topo-Wen,ColloqTopoWen2010} which instead does not
break the symmetry of the system. PTs are also classified into classical
ones and quantum ones, the former are thought to be driven by thermal
fluctuations and the latter by quantum fluctuations~\cite{Sachdev-QPT,Irish2017}. Since the symmetry
requirement of these two classes of PTs are contrary, they are in principle
incompatible. An exceptional finding of their coexistence would be
surprising and intriguing.

When PTs traditionally occur in thermodynamical systems, few-body systems
can also manifest PTs, as it has been found in light-matter interactions.
Indeed, the quantum Rabi model (QRM)~\cite{rabi1936,Rabi-Braak,Eckle-Book-Models},
known as a most fundamental model of light-matter interactions, possesses
a QPT~\cite{Ashhab2013,Ying2015,Hwang2015PRL} in low-frequency limit,
$\omega /\Omega \rightarrow 0$ for the ratio of the bosonic mode frequency $%
\omega $ and the atomic level splitting $\Omega $, which is a replacement of
thermodynamical limit in many-body systems. Although it might be a matter of
taste to term the transition quantum or not by considering the negligible
quantum fluctuations in the photon vacuum state~\cite{Irish2017}, the
transition is found to have the scaling behavior which forms critical
universality as traditional QPTs, such critical universality is not only valid
for anisotropy~\cite{LiuM2017PRL,Ying-2021-AQT} but also holds for the Stark non-linear
coupling~\cite{Ying-Stark-top} and the critical exponents can be
bridged to the thermodynamical case~\cite{LiuM2017PRL}. On the other hand,
apart from the various patterns of explicit~\cite{Ying2020-nonlinear-bias}
or hidden~\cite{Ying-2021-AQT} symmetry breaking as in Landau class of PTs,
the symmetry-protected TPTs also emerge~\cite{Ying-2021-AQT,Ying-gapped-top,Ying-Stark-top,Ying-Spin-Winding}
in these single-qubit systems. Interestingly, these TPTs not only occur at gap
closing~\cite{Ying-2021-AQT,Ying-gapped-top,Ying-Stark-top,Ying-Spin-Winding} as in the conventional TPTs in condensed matter~\cite{Topo-Wen,Hasan2010-RMP-topo,Yu2010ScienceHall,Chen2019GapClosing,TopCriterion,Top-Guan,TopNori}, but also happen
in gapped situations~\cite{Ying-gapped-top,Ying-Stark-top,Ying-Spin-Winding} analogously to the unconventional TPTs in the quantum
spin Hall effect with strong electron-electron interactions~\cite%
{Amaricci-2015-no-gap-closing} and the quantum anomalous Hall effect with
disorder~\cite{Xie-QAH-2021}. The study extension of topological
transitions to excited states in the presence of level anti-crossing also
reveals other unconventional types of TPTs with unmatched wave-function
nodes and spin-winding numbers, as well as topological transitions of spin
knots~\cite{Ying-Spin-Winding}. However, these studies on the single-qubit
TPTs are based on numerical analysis, while a more convincing analytical study is lacking. In
such a situation, the problem of winding angle at infinity remains elusive and the unconventional TPTs
are calling for a clearer understanding~\cite{Ying-Spin-Winding}.

In this work, we present a rigorous study for topological transitions in a
fundamental model of light-matter interactions generally including the
Jaynes-Cummings (JC) linear coupling~\cite{JC-model,JC-Larson2021} and Stark
non-linear coupling~\cite%
{Eckle-2017JPA,Maciejewski-Stark,Ying-Stark-top,Xie2019-Stark} (JC-Stark
model). As the eigen states are composed of two Hermite polynomials, we
rigorously demonstrate that the topological structure of wave function has
an exact correspondence to the spin winding by nodes, and the spin winds
without anti-winding nodes. We also analytically show that the spurious
fractional contribution of the winding angle at infinity to the winding
number is actually integer. Thus, the PTs in the model have a nature of TPTs
and the excitation number is endowed a connotation of topological quantum
number. We also point out that the principal transition is both
symmetry-breaking Landau class of transition and symmetry-protected
topological class of transition simultaneously, while conventionally these
two classes of transitions are incompatible due to the contrary symmetry
requirements. Our results may provide a deeper insight for the few-body
phase transitions in light-matter interactions, including the origin of
unconventional topological transitions.

The paper is organized as follows. Section \ref{Sect-Model} introduces the
JC-Stark model for analytical analysis in this work. Anisotropy is also
included for a further discussion. Section \ref{Sect-Solution} presents the
exact solution of the JC-Stark model. Section \ref{Sect-TopoTranst} shows
the topological nature of the transitions by analytical analysis on the
nodes of eigen wave functions and correspondence of spin windings. Section %
\ref{Sect-Landau-Topo} demonstrates that the principal transition is
simultaneously both Landau class and topological class of transitions.
Section \ref{Sect-Unconv-TPTs} shows the TPTs transition without parity
variation and gives an understanding for the unconventional TPTs without gap
closing for anisotropic case.
Section \ref{Sect-Conclusions} is devoted to conclusions and discussions.

\section{Model and symmetry}
\label{Sect-Model}

We start with a fundamental model of light-matter interactions with Hamiltonian~\cite%
{Ying-Stark-top,Stark-Grimsmo2013}
\begin{eqnarray}
H &=&H_{0}+H_{g}+H_{\lambda }+H_{{\rm Stark}}, \\
H_{0} &=&\omega a^{\dagger }a+\frac{\Omega }{2}\sigma _{x},\quad H_{{\rm %
Stark}}=\chi \omega \hat{n}\sigma _{x}, \\
H_{g} &=&g\left( \widetilde{\sigma }_{-}a^{\dagger }+\widetilde{\sigma }%
_{+}a\right) ,\quad H_{\lambda }=\lambda g\left( \widetilde{\sigma }%
_{+}a^{\dagger }+\widetilde{\sigma }_{-}a\right) , \label{H-Lambda}
\end{eqnarray}%
which generally includes a bosonic mode with photon number $\hat{n}%
=a^{\dagger }a$ and frequency $\omega $, a qubit represented by the Pauli
matrices $\sigma _{x,y,z}$ with level splitting $\Omega $, the rotating-wave
term of interaction $H_{g}$ with coupling strength $g$, the counter-rotating
term $H_{\lambda }$ with coupling anisotropy ratio $\lambda $, and the Stark
non-linear interaction~\cite{Eckle-2017JPA,Maciejewski-Stark,Xie2019-Stark} $%
H_{{\rm Stark}}$ with coupling ratio $\chi $.

In the literature, $H_{{\rm JC}%
}=H_{0}+H_{g}$ is the Jaynes-Cummings model (JCM)~\cite%
{JC-model,JC-Larson2021}, $H_{0}+H_{g}+H_{\lambda }$ is the anisotropic QRM~%
\cite{PRX-Xie-Anistropy} and $\lambda =1$ case is the QRM~\cite%
{rabi1936,Rabi-Braak,Eckle-Book-Models}. Here we define $\widetilde{\sigma }^{\pm
}=(\sigma _{z}\mp i\sigma _{y})/2$ under adoption of spin notation as in Ref.%
\cite{Irish2014}, in which $\sigma _{z}=\pm $ conveniently represents the
two flux states in the flux-qubit circuit system\cite{flux-qubit-Mooij-1999}.
Numerical studies show that these models manifest single-qubit TPTs \cite%
{Ying-2021-AQT,Ying-gapped-top,Ying-Stark-top,Ying-Spin-Winding}. In the
present work, for an analytical analysis we shall first focus on the
JC-Stark model \cite{Ying-Stark-top}
\begin{equation}
H_{{\rm JC-Stark}}=H_{0}+H_{g}+H_{{\rm Stark}},  \label{H-JC-Stark}
\end{equation}%
while in the end we will also use the analytical results to discuss the
unconventional TPTs in the general model $H$. All these model have the parity symmetry, $[\hat{%
P},H]=0$ with $\hat{P}=\sigma _{x}(-1)^{a^{\dagger }a}$, which as we will see
is the key symmetry that protects the TPTs.

To extract the topological feature we rewrite the Hamiltonian in position
space%
\begin{eqnarray}
H &=&\frac{\omega }{2}\hat{p}^{2}+v_{\sigma _{z}}\left( x\right)
+H_{+}\sigma ^{+}+H_{-}\sigma ^{-}, \\
H_{\pm } &=&\frac{\left( \Omega -\chi \omega \right) }{2}\mp g_{y}i\sqrt{2}%
\hat{p}+\frac{\chi \omega }{2}\left( \hat{x}^{2}+\hat{p}^{2}\right) .
\end{eqnarray}%
by transformation $a^{\dagger }=(\hat{x}-i\hat{p})/\sqrt{2},$ $a=(\hat{x}+i%
\hat{p})/\sqrt{2}$, where $\hat{p}=-i\frac{\partial }{\partial x}$, and spin
raising and lowering on $\sigma _{z}=\pm $ basis, $\sigma _{x}=\sigma
^{+}+\sigma ^{-}$, $\sigma _{y}=-i(\sigma _{+}-\sigma _{-})$. In such a
representation $v_{\sigma _{z}}(x)=\omega \left( x+g_{z}^{\prime }\sigma
_{z}\right) ^{2}/2+\varepsilon _{0}^{z}$ is an effective spin-dependent
potential with $g_{z}^{\prime }=\sqrt{2}g_{z}/\omega $, $g_{z}=\frac{\left(
1+\lambda \right) }{2}g$ and $\varepsilon _{0}^{z}=-\frac{1}{2}%
[g_{z}^{\prime 2}+1]\omega $. The $\Omega $ term now acts as spin flipping
in $\sigma _{z}$ space or tunneling in position space \cite%
{Ying2015,Irish2014}. We have also defined $g_{y}=\frac{\left( 1-\lambda
\right) }{2}g$. The $g_{y}$ terms, together as $\sqrt{2}g_{y}\hat{p}\sigma
_{y},$ resemble~\cite{Ying-Stark-top} the Rashba spin-orbit coupling in
nanowires~\cite%
{Nagasawa2013Rings,Ying2016Ellipse,Ying2017EllipseSC,Ying2020PRR,Gentile2022NatElec} or the equal-weight
mixture~\cite{LinRashbaBECExp2011,LinRashbaBECExp2013Review,Ying-gapped-top} of the
linear Dresselhaus~\cite{Dresselhaus1955} and Rashba~\cite{Rashba1984}
spin-orbit couplings.

\section{Exact Solution}

\label{Sect-Solution}

The JC-Stark model \eqref{H-JC-Stark} possesses $U(1)$ symmetry, as
denoted by the excitation number $\hat{n}+\left\vert \Uparrow
\right\rangle \left\langle \Uparrow \right\vert $ or $\hat{n}+\sigma
_{x}/2+1/2$, the eigenstates only involve bases with a same excitation number
$n$ and finally take the following form~\cite{Ying-Stark-top} similar to the JCM~\cite{JC-model,JC-Larson2021}
\begin{eqnarray}
\psi _{n}^{\left( x,\eta \right) } &=&\left( C_{n\Uparrow }^{\left( \eta
\right) }\left\vert n-1,\Uparrow \right\rangle +C_{n\Downarrow }^{\left(
\eta \right) }\left\vert n,\Downarrow \right\rangle \right) /\sqrt{N_{n}},
\label{WaveF-JC-n} \\
\psi _{0} &=&\left\vert 0,\Downarrow \right\rangle
\label{WaveF-JC-0}
\end{eqnarray}%
where $\eta =\pm $ denotes two branches of energy levels, $n=1,2,\cdots $
labels the Fock state on photon-number basis and $\Uparrow ,\Downarrow $ are
two spins states of $\sigma _{x}$. The parity is negative (positive) when $n
$ is even (odd):
\begin{equation}
\hat{P}\psi _{n}^{\left( x,\eta \right) }=\left( -1\right) ^{n-1}\psi
_{n}^{\left( x,\eta \right) },\quad \hat{P}\psi _{0}=\left( -1\right) \psi
_{0}.  \label{Eigen-Parity}
\end{equation}%
The coefficients are explicitly given by%
\begin{eqnarray}
C_{n\Uparrow }^{\left( \eta \right) } &=&e_{-}+\eta \sqrt{e_{-}^{2}+n\ g^{2}}%
,  \label{Cx-up} \\
C_{n\Downarrow }^{\left( \eta \right) } &=&g\sqrt{n},  \label{Cx-down}
\end{eqnarray}%
where $e_{+}=\left( n-\frac{1+\chi }{2}\right) \omega $, $e_{-}=\frac{1}{2}%
\left( \Omega -\omega \right) +\left( n-\frac{1}{2}\right) \chi \omega $ and
$N_{n}=C_{n\Uparrow }^{\left( \eta \right) 2}+C_{n\Downarrow }^{\left( \eta
\right) 2}$ is the normalization factor. For state $\psi _{0}$ one can
define $C_{0\Downarrow }=1$ and $C_{0\Uparrow }=0$ as similar coefficient
notation. Correspondingly the eigenenergies are determined by%
\begin{eqnarray}
E^{\left( n,\eta \right) } &=&e_{+}+\eta \sqrt{e_{-}^{2}+n\ g^{2}},
\label{E-n-JC} \\
E^{0} &=&-\frac{\Omega }{2}.
\end{eqnarray}%
Apparently the energy branch $E^{\left( n,+\right) }$ is higher than $%
E^{\left( n,-\right) }$, thus the ground state is the lowest state of $\psi
_{n}^{\left( x,-\right) }$ and $\psi _{0}$. So far $n$ is only the excitation number and we have not seen any topological aspect.

\begin{figure*}[t]
\includegraphics[width=2.0%
\columnwidth]{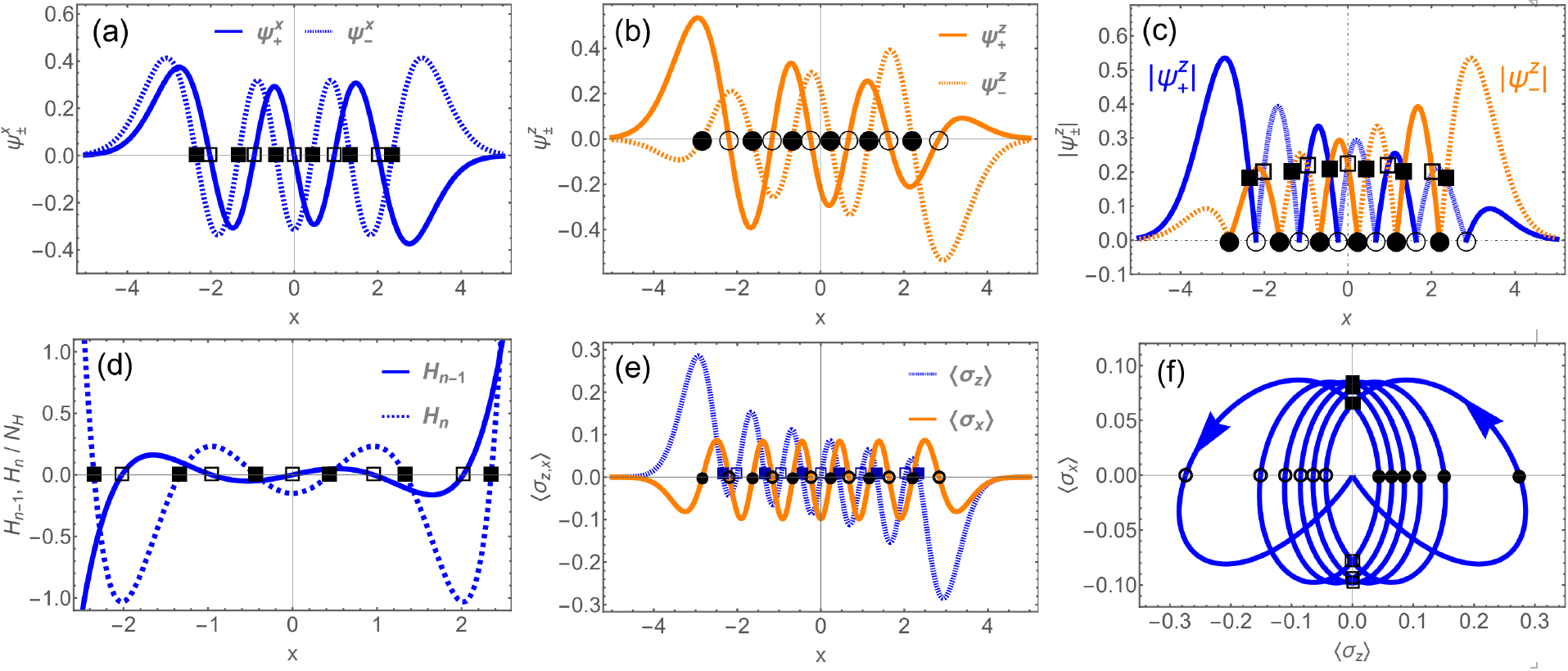}
\caption{ {\it Analytical correspondence of nodes or zeros in wave function components (a-c), Hermite
polynomials (d) and spin windings (e,f).}
The filled (empty) squares mark the corresponding nodes in $\psi^x _{+}$ ($\psi^x _{-}$), $\langle \sigma _{z}\left( x\right) \rangle$ with positive (negative) $\langle \sigma _{x}\left( x\right) \rangle$, $H_{n}$ ($H_{n-1}$) respectively,
while dots (circles) denote the nodes in $\psi^z _{+}$ ($\psi^z _{-}$), $\langle \sigma _{x}\left( x\right) \rangle$ with positive (negative) $\langle \sigma _{z}\left( x\right) \rangle$. Nodes of $\psi^x _{\pm}$ ($\psi^z _{\pm}$) are also the amplitude-crossing points $|\psi^z _{+}|=|\psi^z _{-}|$ ($|\psi^x _{+}|=|\psi^x _{-}|$) as in (c).
Here $\omega=0.5\Omega$, $g=1.5g_{{\rm s}}$, $\lambda=0.2$, $j_E=5$ and $g_{\rm s}=\sqrt{\omega \Omega}/2$.
}
\label{fig-WaveF-SzSx}
\end{figure*}

\begin{figure}[t]
\includegraphics[width=1.0\columnwidth]{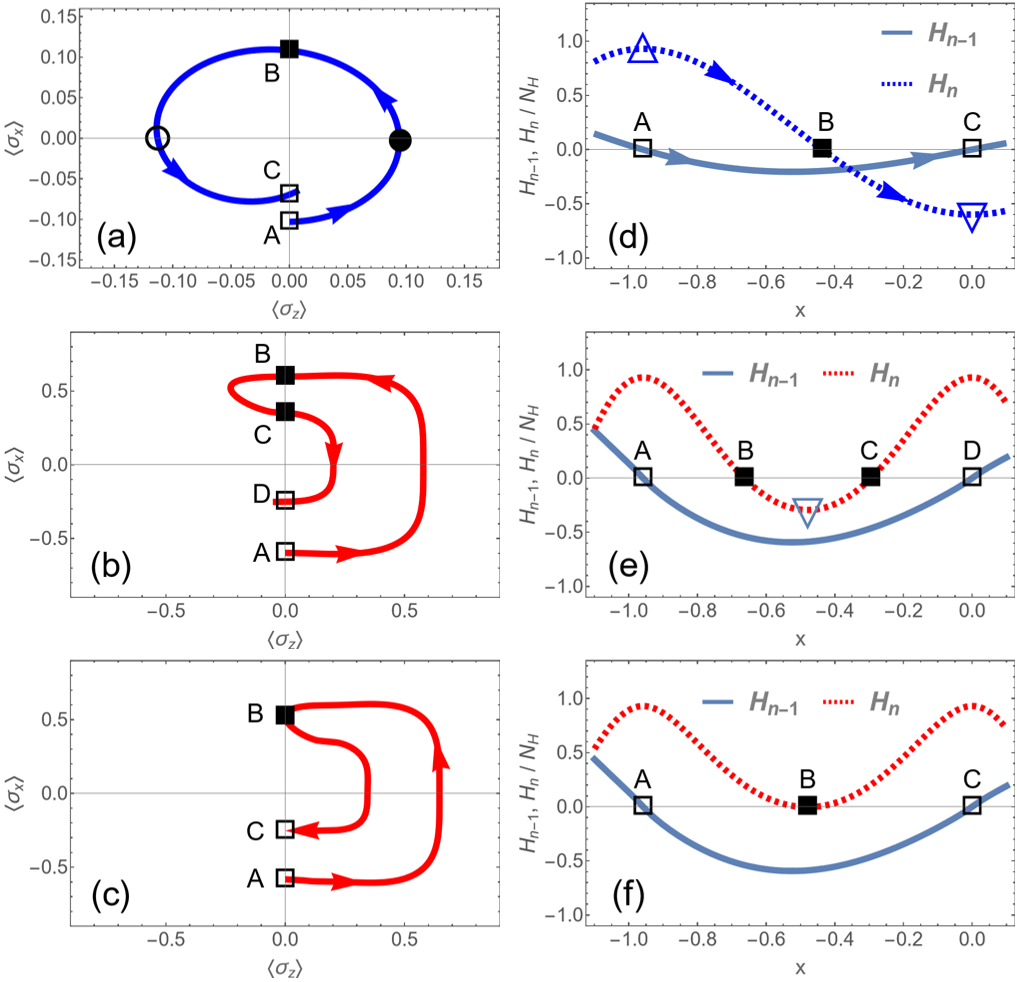}
\caption{{\it Full winding without anti-winding nodes.}
(a-c)
Schematic spin winding between two adjacent nodes (empty squares) on negative $\langle \sigma _x(x) \rangle$ axis, respectively without
anti-winding nodes (a), with two anti-winding nodes (filled squares) on positive $\langle \sigma _x(x) \rangle$ axis (b), and with one anti-winding nodes (c).
(d-f)
Required evolution of Hermite polynomial $H_n(x)$ (dotted lines) for two adjacent roots (empty squares) in $H_{n-1}(x)$, corresponding to (a-c), with (d) fulfilled and (e,f) unfulfilled.
}
\label{fig-Full-winding}
\end{figure}

\begin{figure}[t]
\includegraphics[width=1.0\columnwidth]{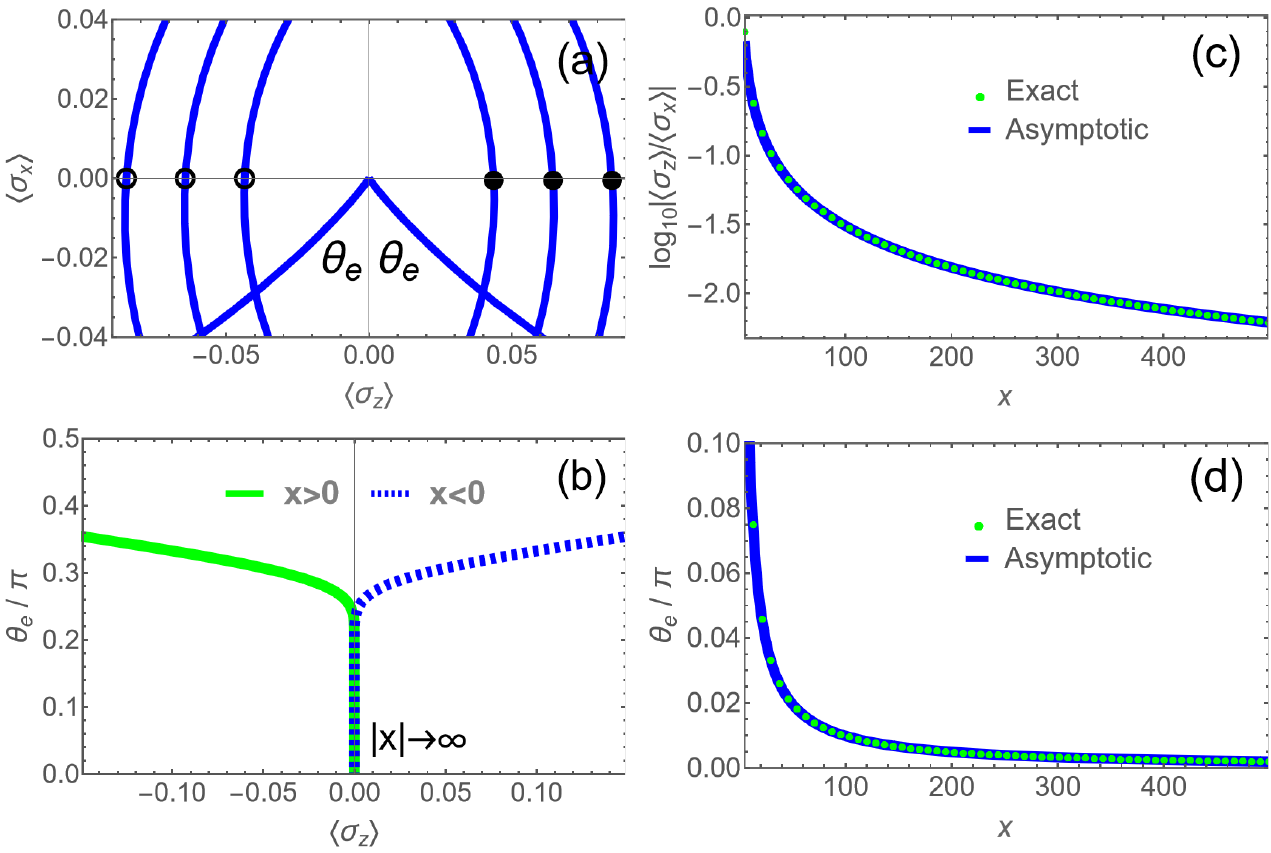}
\caption{{\it Vanishing external spin-winding angle at infinity.}
(a)
A zoom-in plot of the spin winding in Fig.\ref{fig-WaveF-SzSx}(f) around
the origin. $\theta_e$ is the external winding angle at large $x$.
(b)
$\theta_e$ versus $\langle \sigma _x(x) \rangle$ in $|x|\rightarrow \infty$ limit.
(c,d)
$\log _{10}(\langle \sigma _z(x)\rangle / \langle \sigma _x(x) \rangle)$ (c) and $\theta_e$ (d)
versus large $x$ by exact solution (dots) and asymptotic expression \eqref{Eq-Angle-infinity} (solid line).
}
\label{fig-Angle-Infinity}
\end{figure}

\section{Topological-transition nature at finite frequencies}

\label{Sect-TopoTranst}

\subsection{Wave-function nodes}

We can rewrite the eigen wave function in position space
\begin{eqnarray}
\psi _{+}^{x}\left( x\right) &\equiv &\psi _{n,\Uparrow }^{\left( x,\eta
\right) }\left( x\right) =C_{n\Uparrow }^{\left( \eta \right) }\phi
_{n-1}\left( x\right) /\sqrt{N_{n}},  \label{wave-X-1} \\
\psi _{-}^{x}\left( x\right) &\equiv &\psi _{n,\Downarrow }^{\left( x,\eta
\right) }\left( x\right) =C_{n\Downarrow }^{\left( \eta \right) }\phi
_{n}\left( x\right) /\sqrt{N_{n}}.  \label{wave-X-2}
\end{eqnarray}%
where $\phi _{n}\left( x\right) $ is the eigenstate of quantum harmonic
oscillator with quantum number $n$%
\begin{equation}
\phi _{n}\left( x\right) =\frac{1}{\pi ^{1/4}\sqrt{2^{n}n!}}H_{n}\left(
x\right) e^{-x^{2}/2}.
\end{equation}%
Note the Hermite polynomial $H_{n}\left( x\right) $ has an $n$ number of
real roots, $x=y_{{\rm Z}}$ where $H_{n}\left( y_{{\rm Z}}\right) =0,$
accordingly the wave function components $\psi _{n,\Uparrow }^{\left( x,\eta
\right) }\left( x\right) $ and $\psi _{n,\Downarrow }^{\left( x,\eta \right)
}\left( x\right) $ respectively have $n-1$ and $n$ numbers of real nodes $y_{%
{\rm Z}}$ where $\psi _{n,\sigma _{x}}^{\left( x,\pm \right) }\left( y_{{\rm %
Z}}\right) =0$.

We can also transform onto the spin-$\sigma _{z}$ basis, represented by $%
\uparrow $ and $\downarrow $, on which the wave function becomes
\begin{equation}
\psi _{n}^{\left( z,\eta \right) }=\ \psi _{+}^{z}\left( x\right) \left\vert
\uparrow \right\rangle +\psi _{-}^{z}\left( x\right) \left\vert \downarrow
\right\rangle  \label{wave-Z-0}
\end{equation}%
with spin components%
\begin{eqnarray}
\psi _{+}^{z}\left( x\right) &\equiv &\psi _{n,\uparrow }^{\left( z,\eta
\right) }\left( x\right) =\frac{C_{n\Uparrow }^{\left( \eta \right) }\phi
_{n-1}\left( x\right) +C_{n\Downarrow }^{\left( \eta \right) }\phi
_{n}\left( x\right) }{\sqrt{2N_{n}}},  \label{wave-Z-1} \\
\psi _{-}^{z}\left( x\right) &\equiv &\psi _{n,\downarrow }^{\left( z,\eta
\right) }\left( x\right) =\frac{C_{n\Uparrow }^{\left( \eta \right) }\phi
_{n-1}\left( x\right) -C_{n\Downarrow }^{\left( \eta \right) }\phi
_{n}\left( x\right) }{\sqrt{2N_{n}}}.  \label{wave-Z-2}
\end{eqnarray}%
The parity symmetry ensures
\begin{equation}
\psi _{n,\uparrow }^{\left( z,\eta \right) }\left( x\right) =\left(
-1\right) ^{n-1}\psi _{n,\downarrow }^{\left( z,\eta \right) }\left(
-x\right).  \label{wave-Z-Inverse-x}
\end{equation}%
Later on in Sect. \ref{Sect-Full-Winding} we will see that both components have $%
n$ number of nodes, $x=x_{{\rm Z}}$, where $\psi _{n,\sigma _{z}}^{\left(
z,\eta \right) }\left( x_{{\rm Z}}\right) =0.$

We give an example of eigen wave function in Fig. \ref{fig-WaveF-SzSx} (a-c)
for $n=6$ with $\psi _{\pm }^{x,z}\left( x\right) \ $representing $\psi
_{n,\sigma _{x}}^{\left( x,\eta \right) }\left( x\right) $ and $\psi
_{n,\sigma _{z}}^{\left( z,\eta \right) }\left( x\right) $ in Eqs. (\ref{wave-X-1}%
,\ref{wave-X-2},\ref{wave-Z-1},\ref{wave-Z-2}). Here in the figure the nodes of $\psi
_{\pm }^{x}\left( x\right) $ are marked by empty boxes (spin-up, $+=\Uparrow
$) and filled boxes (spin-down, $-=\Downarrow $) in panel (a), and the nodes
of $\psi _{\pm }^{z}\left( x\right) $ by circles (spin-up, $+=\uparrow $)
and dots (spin-down, $-=\downarrow $) in panel (b). We can also plot all the
nodes together in the wave-function amplitude $\left\vert \psi _{\pm }^{z}\left( x\right) \right\vert $
as in panel (c) where the nodes of $\psi _{n,\sigma _{x}}^{\left( x,\eta
\right) }\left( x\right) $ are located at points $\left\vert \psi
_{+}^{z}\left( x\right) \right\vert =\left\vert \psi _{-}^{z}\left( x\right)
\right\vert $. Here the empty boxes ($\Uparrow$) are crossing points of $\psi
_{+}^{z}\left( x\right) =\psi _{-}^{z}\left( x\right) $ in both solid lines ($\psi _{\pm}^{z} >0$)
or both broken lines ($\psi _{\pm}^{z} <0$), while filled boxes ($\Downarrow $) are crossing points of
$\psi _{+}^{z}\left( x\right) =-\psi _{-}^{z}\left( x\right) $ between solid
lines and broken lines.

The node number represents the topological structure of the wave function in
the sense that with a fixed node number one cannot go to another node state
by continuous shape deformation of the wave function, just as one cannot
change a torus into a sphere by a continuous deformation in the topological
picture of so-called rubber-sheet geometry. Such wave-function topological
structure can be reflected by physical topological character as there is a
one-to-one correspondence between the wave-function nodes and the spin-winding
nodes, as we shall discuss in the following sections.

\subsection{Spin winding: Node correspondence to wave function and
symmetric/anti-symmetric properties}

Note the eigen wave functions in \eqref{WaveF-JC-n} and \eqref{WaveF-JC-0} are real, so that the corresponding spin texture are
related to the wave function components by
\begin{eqnarray}
\langle \sigma _{z}\left( x\right) \rangle &=&\psi _{+}^{z}\left( x\right)
^{2}-\psi _{-}^{z}\left( x\right) ^{2}=2\psi _{+}^{x}\left( x\right) \psi
_{-}^{x}\left( x\right)  \label{SpinZ-byWave} \\
\langle \sigma _{x}\left( x\right) \rangle &=&\psi _{+}^{x}\left( x\right)
^{2}-\psi _{-}^{x}\left( x\right) ^{2}=2\psi _{+}^{z}\left( x\right) \psi
_{-}^{z}\left( x\right) ,  \label{SpinX-byWave} \\
\langle \sigma _{y}\left( x\right) \rangle &=&i\left[ \psi _{-}^{z}\left(
x\right) ^{\ast }\psi _{+}^{z}\left( x\right) -\psi _{+}^{z}\left( x\right)
^{\ast }\psi _{-}^{z}\left( x\right) \right] =0.  \label{SpinY-byWave}
\end{eqnarray}%
We see the spin are winding within $\langle \sigma _{z}\left( x\right)
\rangle $-$\langle \sigma _{z}\left( x\right) \rangle $ plane, and the nodes
of eigen wave function are in one-to-one correspondence to the nodes of the
spin winding:%
\begin{eqnarray}
\psi _{n,\sigma _{z}}^{\left( z,\eta \right) }\left( x_{{\rm Z}}\right)
&=&0\Longleftrightarrow \langle \sigma _{z}\left( x_{{\rm Z}}\right) \rangle
=0, \\
\psi _{n,\sigma _{x}}^{\left( x,\eta \right) }\left( y_{{\rm Z}}\right)
&=&0\Longleftrightarrow \langle \sigma _{x}\left( y_{{\rm Z}}\right) \rangle
=0.
\end{eqnarray}
The node correspondence of the wave function and the spin winding is shown by panels (a-c) and (e,f) in Fig. \ref{fig-WaveF-SzSx},
where the squares represent the corresponding nodes of $\psi^x _{\pm}$ and $\langle \sigma _{z}\left( x\right) \rangle$ and the dots or circles locate the corresponding nodes of $\psi^z _{\pm}$ and $\langle \sigma _{x}\left( x\right) \rangle$.

From Eqs. (\ref{SpinZ-byWave}-\ref{SpinY-byWave}) the spin texture for state $\psi _{n}^{\left( x,\eta \right) } $ can be
analytically obtained to be%
\begin{eqnarray}
&&\langle \sigma _{z}\left( x\right) \rangle =\frac{e^{-x^{2}}g\ C_{n\omega
\eta }}{2^{n-3/2}N_{\sigma }}H_{n-1}\left( x\right) H_{n}\left( x\right) ,
\label{SpinZ-expression} \\
&&\langle \sigma _{x}\left( x\right) \rangle =\frac{e^{-x^{2}}}{%
2^{n}N_{\sigma }}[C_{n\omega \eta }^{2}H_{n-1}\left( x\right)
^{2}-2g^{2}H_{n}\left( x\right) ^{2}],  \label{SpinX-expression} \\
&&\langle \sigma _{y}\left( x\right) \rangle =0,
\end{eqnarray}%
where $N_{\sigma }=\sqrt{\pi }(n-1)!\ [4g^{2}n+C_{n\omega \eta }\Omega
_{n\omega \chi }]$, $C_{n\omega \eta }=\Omega _{n\omega \chi }+\eta \sqrt{%
\Omega _{n\omega \chi }^{2}+4g^{2}n}$, $\Omega _{n\omega \chi }=[\Omega
-\omega +\left( 2n-1\right) \chi \omega ]$. For state $\psi _{0}$, we have $\langle \sigma _{z}\left( x\right) \rangle=\langle \sigma _{y}\left( x\right) \rangle=0$ and $\langle \sigma _{x}\left( x\right) \rangle= - e^{-x^{2}}/\sqrt{\pi} $.  Eqs. \eqref{SpinZ-expression} and \eqref{SpinX-expression} indicate that there is also a correspondence of the roots of the Hermite polynomials to the nodes of the wave function and the spin winding, as illustrated by Fig. \ref{fig-WaveF-SzSx}(d). We will leave more discussions around Eq.~\eqref{Eq-H-roots} in Sect.~\ref{Sect-invariant-nodes} and with $f_{\pm}$ in Sect.~\ref{Sect-Full-Winding}.

The parity symmetry also leads to the symmetry of $\langle \sigma _{x}\left(
x\right) \rangle $ and anti-symmetry of $\langle \sigma _{z}\left( x\right)
\rangle $. Indeed, the parity symmetry implies $\psi _{-}^{z}\left( x\right)
=P\psi _{+}^{z}\left( -x\right) $ \cite{Ying2020-nonlinear-bias},
substitution of which into \eqref{SpinZ-byWave} and \eqref{SpinX-byWave}
yields
\begin{equation}
\langle \sigma _{z}\left( -x\right) \rangle =-\langle \sigma _{z}\left(
x\right) \rangle ,\quad \langle \sigma _{x}\left( -x\right) \rangle
=+\langle \sigma _{x}\left( x\right) \rangle .  \label{spin-symmetry}
\end{equation}%
The above symmetric and anti-symmetric properties of $\langle \sigma
_{x}\left( x\right) \rangle $ and $\langle \sigma _{z}\left( x\right)
\rangle $ can also be directly seen from \eqref{SpinZ-expression} and %
\eqref{SpinX-expression} as $H_{n}\left( -x\right) =\left( -1\right)
^{n}H_{n}\left( x\right) $. Fig. \ref{fig-WaveF-SzSx}(e) shows an example of
the spin texture, one sees that indeed $\langle \sigma _{x}\left( x\right)
\rangle $ is symmetric and $\langle \sigma _{z}\left( x\right) \rangle $ is
anti-symmetric, which yields a $\langle \sigma _{z}\left( x\right) \rangle $%
-reflection-symmetric spin winding in the $\langle \sigma _{z}\left(
x\right) \rangle $-$\langle \sigma _{x}\left( x\right) \rangle $ plane as
demonstrated in Fig. \ref{fig-WaveF-SzSx}(f). These symmetry properties of
the spin texture will be used in the argument for the distribution of $%
\langle \sigma _{x}\left( x\right) \rangle $ nodes.

\subsection{Invariant $\langle \sigma _{z}\left( x\right) \rangle $
nodes}
\label{Sect-invariant-nodes}

From Eq. \eqref{SpinZ-expression} we see the nodes of $\langle \sigma
_{z}\left( x\right) \rangle $ are completely decided by the roots of $%
H_{n-1}\left( x\right) $ and $H_{n}\left( x\right) $. The $\langle \sigma
_{z}\left( x\right) \rangle $ nodes are located at the roots of the Hermite
polynomials%
\begin{equation}
H_{n-1}(z_{{\rm Z}}^{\left( n-1\right) })=0,\text{ or }H_{n}(z_{{\rm Z}%
}^{\left( n\right) })=0,
\label{Eq-H-roots}
\end{equation}%
which are independent of the model parameters. Such an invariant feature may
provide some particular advantage in designing potential topological
devices. For an example, these spin nodes could provide a topological
information for quantum topological encoding and decoding \cite%
{Ying-Spin-Winding}, the topological information based on such a kind of
invariant nodes will be robust as it is completely unaffected by the
variations of the parameters within the topological phase.

\subsection{Full winding without anti-winding nodes}
\label{Sect-Full-Winding}

It should be noted that the Hermite-polynomial roots $z_{{\rm Z}}^{\left( n-1\right) },z_{{\rm Z}%
}^{\left( n\right) }$ are alternate due to the relation
\begin{equation}
H_{n}^{\prime }\left( x\right) \equiv \partial _{x}H_{n}\left( x\right)
=2nH_{n-1}\left( x\right)  \label{dHn=Hn}
\end{equation}%
which indicates that the $H_{n-1}$ roots, $z_{Z}^{\left( n-1\right) },$ are
always the maxima or minima of $H_{n}\left( x\right) $, as shown by the upward and downward triangles in Fig. \ref%
{fig-Full-winding}(d). And from relation $H_{n}\left( x\right)
=2xH_{n-1}\left( x\right) -\partial _{x}H_{n-1}\left( x\right) $ we see%
\begin{equation}
H_{n}(z_{{\rm Z}}^{\left( n-1\right) })=-H_{n-1}^{\prime }(z_{{\rm Z}%
}^{\left( n-1\right) })
\end{equation}%
which indicates that two adjacent roots $z_{Z,j}^{\left( n-1\right)
},z_{Z,j+1}^{\left( n-1\right) }$ must have different signs of $H_{n}$ due
to the different gradient signs from $H_{n-1}^{\prime }$, i.e.
\begin{equation}
H_{n}(z_{{\rm Z},j}^{\left( n-1\right) })H_{n}(z_{{\rm Z},j+1}^{\left(
n-1\right) })<0,
\end{equation}%
leading to a root $z_{Z}^{\left( n\right) }$ between $z_{Z,j}^{\left(
n-1\right) }$ and $z_{Z,j+1}^{\left( n-1\right) }$ (Note here $H_{n}(z_{{\rm %
Z},j}^{\left( n-1\right) })$, $H_{n}(z_{{\rm Z},j+1}^{\left( n-1\right) })$
are always finite due to the Tur\'{a}n inequality $H_{n}\left( x\right)
^{2}-H_{n-1}\left( x\right) H_{n+1}\left( x\right) >0$ which excludes
simultaneous zeros of $H_{n}\left( x\right) $ and $H_{n-1}\left( x\right) $%
). Fig. \ref{fig-Full-winding}(b) shows such a case schematically: the empty
boxes A and C represent the two adjacent $H_{n-1}$ roots, at which the
gradient has different signs as indicated by the arrow orientations along
the solid line. The $H_{n}$ signs at A and C are different (indicated by the
triangles) which must surround an $H_{n}$ root (filled square B).

Note from Eqs.~\eqref{SpinZ-expression} and \eqref{SpinX-expression} we find $H_{n-1}$ roots and $H_{n}$ roots
correspond to $\langle \sigma _{z}\left( x\right) \rangle $ nodes on
opposite $\langle \sigma _{x}\left( x\right) \rangle $ axes, as in Fig. \ref%
{fig-Full-winding}(a). Thus, the above analysis means that between two adjacent $%
\langle \sigma _{z}\left( x\right) \rangle $ nodes on same $\langle \sigma
_{x}\left( x\right) \rangle $ axis (empty squares A,C) there must be another
$\langle \sigma _{z}\left( x\right) \rangle $ node on the opposite $\langle
\sigma _{x}\left( x\right) \rangle $ axis (filled square B). The possibility
to have more than one $\langle \sigma _{z}\left( x\right) \rangle $ nodes on
the opposite $\langle \sigma _{x}\left( x\right) \rangle $ axis as in Fig. %
\ref{fig-Full-winding}(b) is excluded, as that would spuriously bring some
new $H_{n}$ maximum or minimum which however has no match of $H_{n-1}$ zero,
violating relation \eqref{dHn=Hn}, as denoted by the triangle in Fig. \ref%
{fig-Full-winding}(e). This excluded case in Fig. \ref{fig-Full-winding}(b)
also avoids anti-winding (i.e. cancelation of spin-winding angle in route returning). The
anti-winding at one $\langle \sigma _{z}\left( x\right) \rangle $ node as in Fig. \ref{fig-Full-winding}(c)
is also violating relation \eqref{dHn=Hn}. Therefore,
the $\sigma _{z}\left( x\right) \rangle $ nodes should appear alternately on
positive and negative $\langle \sigma _{x}\left( x\right) \rangle $ axes
without anti-winding.

The same happens for $\langle \sigma _{x}\left( x\right) \rangle $ nodes.
Actually Eq. \eqref{SpinX-expression} can be factorized into a product of
factors $f_{\pm }=H_{n-1}\left( x\right) \pm cH_{n}\left( x\right) $ where $%
c $ is parameter-decided coefficient. The $\langle \sigma _{x}\left( x\right) \rangle $ nodes are just the $f_{\pm }$ roots.
Both factors $f_{\pm }$ have $n$ roots as $%
H_{n-1}\left( x\right) $ and $H_{n}\left( x\right) $ subject to relation %
\eqref{dHn=Hn} are interlacing with alternate zeros as in Fig. \ref%
{fig-WaveF-SzSx}(d), while the number of the $f_{\pm }$ roots is decided by
the crossing times of $H_{n-1}\left( x\right) $ and $H_{n}\left( x\right) $
which are unaffected by any amplitude amplification with nonzero $c$, as one
can recognize from Fig. \ref{fig-Full-winding}(d). Thus, there are $2n$ of $\langle \sigma _{x}\left( x\right) \rangle $ nodes, while from
the Eq. \eqref{SpinX-expression} we have known there are $2n-1$ of $\langle \sigma _{z}\left( x\right) \rangle $ nodes. Since there is no anti-winding, the $\langle \sigma _{x}\left( x\right) \rangle $ nodes and the $\langle \sigma _{z}\left( x\right) \rangle $ nodes are also interlacing. Indeed, except the node on the infinity side, each $\langle \sigma
_{x}\left( x\right) \rangle $ node can only appear in the interval between two
adjacent $\langle \sigma _{z}\left( x\right) \rangle $ nodes on opposite $%
\langle \sigma _{x}\left( x\right) \rangle $ axes, while each interval can accommodate one and
only one $\langle \sigma _{x}\left( x\right) \rangle $ node, otherwise
accommodation of more $\langle \sigma _{x}\left( x\right) \rangle $ nodes
would totally outnumber the $2n$ number of $\langle \sigma
_{x}\left( x\right) \rangle $ nodes due to the $\langle \sigma _{z}\left(
x\right) \rangle $-reflection-symmetry afore-mentioned at %
\eqref{spin-symmetry}.

Wrapping up the above analysis points one rigorously comes to the conclusion that the
spin is in full winding without anti-winding nodes, the nodes come in counterclockwise sequence 1234 as in Fig. \ref{fig-Full-winding}(a) or clockwise sequence 1432 on: (1) positive-$\langle \sigma _{z}\left( x\right) \rangle $ axis
(dot), (2) positive-$\langle \sigma _{x}\left( x\right) \rangle $ axis
(filled square), (3) negative-$\langle \sigma _{z}\left( x\right) \rangle $
axis (dot), (4) negative-$\langle \sigma _{x}\left( x\right) \rangle $ axis
(empty square), periodically till
completing the total spin winding at infinity. Such clarification of the full winding behavior is necessary for the explicit extraction of the winding number later on in Sect. \ref{Sect-nW-by-nodes}.

\subsection{Spurious fractional winding angle at infinity}

\label{Sect-Angle-Infinity}

Apart from the main part of spin winding in the node sequence, the total
winding angle is also partially decided by the winding at infinity. A focus plot
on the external winding angle $\theta _{e}$ is illustrated in Fig. \ref%
{fig-Angle-Infinity}(a). At a first glance one might think $\theta _{e}$ is
fractionally finite. However a more careful tracking of $\theta _{e}$ at
larger $\left\vert x\right\vert $ reveals that $\theta _{e}$ is approaching
to zero at infinity, as shown in Fig. \ref{fig-Angle-Infinity}(b). Indeed, at
infinity the leading term of Hermite polynomials is
\begin{equation}
H_{n}\left( x\right) \rightarrow \left( 2x\right) ^{n},
\end{equation}%
so that we have the ratio of the spin texture%
\begin{equation}
\frac{\langle \sigma _{z}\left( x\right) \rangle }{\langle \sigma _{x}\left(
x\right) \rangle }\rightarrow -\frac{\Omega _{n\omega \chi }+\eta \sqrt{%
4g^{2}n+\Omega _{n\omega \chi }^{2}}}{\sqrt{2}g}\frac{1}{x},
\label{Eq-Angle-infinity}
\end{equation}%
which is approaching to zero. As demonstrated in Fig. \ref{fig-Angle-Infinity}(c),
this asymptotic behavior (solid line) agrees well with the exact ratio
(dots) obtained by Eqs. \eqref{SpinZ-expression} and \eqref{SpinX-expression}.
Correspondingly, as shown in Fig. \ref{fig-Angle-Infinity}(d), the external
angle of spin winding is vanishing at infinity%
\begin{equation}
\theta _{\infty }=\arctan \frac{\langle \sigma _{z}\left( x\right) \rangle }{
\langle \sigma _{x}\left( x\right) \rangle }\rightarrow 0.
\end{equation}%
This vanishing external angle achieves an integer number of total spin
winding angle as formulated in next section.

\subsection{Winding number in terms of nodes}
\label{Sect-nW-by-nodes}

One can know the rounds of spin winding by the winding number around the
origin in the $\langle \sigma _{z}\rangle $-$\langle \sigma _{x}\rangle $
plane as calculated by
\begin{equation}
n_{zx}=\frac{1}{2\pi }\int_{-\infty }^{\infty }\frac{\langle \sigma
_{z}\left( x\right) \rangle \partial _{x}\langle \sigma _{x}\left( x\right)
\rangle -\langle \sigma _{x}\left( x\right) \rangle \partial _{x}\langle
\sigma _{z}\left( x\right) \rangle }{\langle \sigma _{z}\left( x\right)
\rangle ^{2}+\langle \sigma _{x}\left( x\right) \rangle ^{2}}dx,
\label{n-zx}
\end{equation}
which has also been applied in topological classification in nanowire
systems and quantum systems with geometric driving~\cite%
{Ying2016Ellipse,Ying2017EllipseSC,Ying2020PRR,Gentile2022NatElec}. With the
normalized spin texture%
\begin{equation}
\langle \overline{\sigma }_{z,x}\left( x\right) \rangle =\frac{\langle
\sigma _{z,x}\left( x\right) \rangle }{\sqrt{\langle \sigma _{z}\left(
x\right) \rangle ^{2}+\langle \sigma _{x}\left( x\right) \rangle ^{2}}},
\end{equation}%
we can rewrite the integrand to be%
\begin{eqnarray}
&&\frac{\langle \sigma _{z}\left( x\right) \rangle \partial _{x}\langle
\sigma _{x}\left( x\right) \rangle -\langle \sigma _{x}\left( x\right)
\rangle \partial _{x}\langle \sigma _{z}\left( x\right) \rangle }{\langle
\sigma _{z}\left( x\right) \rangle ^{2}+\langle \sigma _{x}\left( x\right)
\rangle ^{2}}  \nonumber \\
&=&\frac{-\partial _{x}\langle \overline{\sigma }_{z}\left( x\right) \rangle
}{\eta _{x}\sqrt{1-\langle \overline{\sigma }_{z}\left( x\right) \rangle ^{2}
}}=\frac{\partial _{x}\langle \overline{\sigma }_{x}\left( x\right) \rangle
}{\eta _{z}\sqrt{1-\langle \overline{\sigma }_{x}\left( x\right) \rangle ^{2}
}},  \label{Integrand-2}
\end{eqnarray}
where $\eta _{x,z}={\rm sign}\langle \overline{\sigma }_{x,z}\left( x\right)
\rangle $, so that the integral \eqref{n-zx} can be worked out explicitly in
terms of either $\langle \sigma _{x}\left( x\right) \rangle $ or $\langle
\sigma _{z}\left( x\right) \rangle $ nodes
\begin{eqnarray}
n_{zx} &=&-\sum_{i=0}^{M_{x}}\frac{\arcsin \langle \overline{\sigma }
_{z}\left( x_{Z,i+1}\right) \rangle -\arcsin \langle \overline{\sigma }
_{z}\left( x_{Z,i}\right) \rangle }{2\pi \eta _{x}\left( i\right) }
\label{nW-spinZ} \\
&=&\sum_{i=1}^{M_{z}}\frac{\arcsin \langle \overline{\sigma }_{x}\left(
y_{Z,i+1}\right) \rangle -\arcsin \langle \overline{\sigma }_{x}\left(
y_{Z,i}\right) \rangle }{2\pi \eta _{z}\left( i\right) }.  \label{nW-spinX}
\end{eqnarray}
Attention should be paid here that the summation in Eq.\eqref{nW-spinZ} (Eq.
\eqref{nW-spinX}) is respectively over the $M_{x}$ number of $\langle \sigma
_{x}\left( x\right) \rangle $ nodes (the $M_{z}$ number of $\langle \sigma
_{z}\left( x\right) \rangle $ nodes), i.e. over $x_{Z,i}$ ($y_{Z,i}$), not
over the nodes of the variable $\langle \sigma _{z}\left( x\right) \rangle $
($\langle \sigma _{x}\left( x\right) \rangle $) in the integrand %
\eqref{Integrand-2}. Corresponding to\ $\eta _{x,z}$ in \eqref{Integrand-2},
$\eta _{x}\left( i\right) $ ($\eta _{z}\left( i\right) $) is the sign of
$\langle \sigma _{x}\left( x\right) \rangle $ ($\langle \sigma _{z}\left(
x\right) \rangle $) in space section $x\in (x_{Z,i},x_{Z,i+1})$ ($x\in
(y_{Z,i},y_{Z,i+1})$), which can be represented by the sign of a $\langle
\sigma _{z}\left( x\right) \rangle $ node ($\langle \sigma _{x}\left(
x\right) \rangle $ node) in the section. The edge sections $i=0,M_{x,z}$ are
$(-\infty ,x_{Z,1})$ and $(x_{Z,M_{x}},\infty )$ ($(-\infty ,y_{Z,1})$ and
$(y_{Z,M_{z}},\infty )$). We have set $x_{Z,0}=-\infty $ and $%
x_{Z,M_{x}+1}=\infty $ ($y_{Z,0}=-\infty $ and $y_{Z,M_{z}+1}=\infty $).

Note $\arcsin \langle \overline{\sigma }_{z}\left( y_{Z,i}\right) \rangle =
\frac{\pi }{2}{\rm sign}[\overline{\sigma }_{z}\left( y_{Z,i}\right) ]$ and
$\arcsin \langle \overline{\sigma }_{x}\left( x_{Z,i}\right) \rangle =\frac{
\pi }{2}{\rm sign}[\overline{\sigma }_{x}\left( x_{Z,i}\right) ]$, we arrive
at
\begin{eqnarray}
n_{zx} &=&-\sum_{i=0}^{M_{x}}\frac{sgn\langle \overline{\sigma }_{z}\left(
x_{Z,i+1}\right) \rangle -sgn\langle \overline{\sigma }_{z}\left(
x_{Z,i}\right) \rangle }{4\eta _{x}\left( i\right) }  \label{nW-spinZ-Sign}
\\
&=&\sum_{i=0}^{M_{z}}\frac{sgn\langle \overline{\sigma }_{x}\left(
y_{Z,i+1}\right) \rangle -sgn\langle \overline{\sigma }_{x}\left(
y_{Z,i}\right) \rangle }{4\eta _{z}\left( i\right) }  \label{nW-spinX-Sign}
\end{eqnarray}%
where we have set the function $sgn(\overline{\sigma }_{x,z})={\rm sign}(
\overline{\sigma }_{x,z})$ for the nodes and $sgn(\overline{\sigma }
_{x,z})=2\arcsin (\overline{\sigma }_{x,z})/\pi $ for the infinity ends.
Finally only the neighboring nodes with opposite signs have contributions.

Expressions \eqref{nW-spinZ}-\eqref{nW-spinZ-Sign} are valid for general
spin windings. Note the original version of winding number \eqref{n-zx}
involves calculus of both integral and differential, which is numerically
more difficult to treat. In contrast, Eqs. \eqref{nW-spinZ-Sign} and %
\eqref{nW-spinZ-Sign} are simple algebraic expressions comprising only a
finite number of nodes of $\langle \sigma _{z}\left( x\right) \rangle $ and
$\langle \sigma _{x}\left( x\right) \rangle $, which much simplifies the
calculation of the winding number. Moreover, the integral \eqref{n-zx}
depends on the topological structure of the spin texture geometrically, the
equivalence of Eqs. \eqref{nW-spinZ-Sign} and \eqref{nW-spinZ-Sign} to Eq. \eqref{n-zx}
indicates that, given the few points of nodes, the topological
winding number remains the same no matter how the spin texture is
geometrically deformed, which reclaims the original sense of topological
classification in the so-called rubber-sheet geometry. As Eqs. %
\eqref{nW-spinZ-Sign} and \eqref{nW-spinZ-Sign} are the nodes the order of
which encodes the topological message by an algebraic code as in the end of
Sect. \ref{Sect-Full-Winding}, it is also a demonstration of bridging of the
geometrical topology and the algebraic topology, here physically in context
of wave function and spin winding.

According to the discussions in Sections \ref{Sect-Full-Winding} and \ref{Sect-Angle-Infinity},
the spin is in full winding without anti-winding
nodes and the external winding angle at infinity is vanishing. Note there
are $M_{x}=2n$ of $\langle \sigma _{x}\left( x\right) \rangle $ nodes and $
M_{z}=2n-1$ of $\langle \sigma _{z}\left( x\right) \rangle $ nodes, while
the infinity ends only contribute to $sgn(\overline{\sigma }_{z})$ to complete a full integer rounds of winding. Thus,
from \eqref{nW-spinZ} and \eqref{nW-spinX} we can readily conclude that the
magnitude of spin winding number is
\begin{equation}
\left\vert n_{{\rm w}}\right\vert =\left\vert n_{zx}\right\vert =n.
\end{equation}
The sign of $n_{{\rm w}}$ is decided by the winding direction, which is
reflected in $\eta _{x,z}\left( i\right) $ and can be more explicitly
obtained by the status at infinity as in the following.

\subsection{Winding direction}
\label{Sec-nW-direction}

Since the winding is smooth without anti-winding nodes and the external
winding angle is zero at infinity, $\theta _{\infty }\rightarrow 0$, the
winding direction can be determined by the signs of $\langle \sigma
_{z}\left( x\right) \rangle $ and $\langle \sigma _{x}\left( x\right)
\rangle $ at infinity where the spin winding starts and ends. The winding
will be counter-clockwise if $\langle \sigma _{z}\left( x\right) \rangle $
starts to grow negatively (positively) while $\langle \sigma _{x}\left(
x\right) \rangle $ increases positively (negatively), which happens in the
2nd (4th) quadrant; Otherwise the winding is clockwise if $\langle \sigma
_{z}\left( x\right) \rangle $ and $\langle \sigma _{x}\left( x\right)
\rangle $ start in the 1st (3rd) quadrant. Clockwise winding starting in the
2nd (4th) quadrant or counter-clockwise winding starting in the
1st (3rd) quadrant is excluded as that would lead to $M_z=2n+1>M_x=2n$ which conflicts with the
afore-discussed node numbers determined by Eqs.\eqref{SpinZ-expression} and \eqref{SpinX-expression}. Thus, the winding is
counter-clockwise (clockwise) if the sign
\begin{eqnarray}
s_{{\rm w}} &=&\text{sign}\frac{\langle \sigma _{z}\left( x\right) \rangle }{%
\langle \sigma _{x}\left( x\right) \rangle }|_{x\rightarrow -\infty }
\nonumber \\
&=&\text{sign}\left[ \Omega _{n\omega \chi }+\eta \sqrt{4g^{2}n+\Omega
_{n\omega \chi }^{2}}\right]
\end{eqnarray}%
is negative (positive). This indicates that all states with $\eta =-1$ have
a counter-clockwise spin winding direction, while the winding direction of
the states with $\eta =+1$ is opposite. The ground state is composed of
$\eta =-1$ states thus has a counter-clockwise winding direction.

Thus, the energy branch label $\eta $ and the excitation number $n$ together
give the complete information of the spin winding number for state $\psi
_{n}^{\left( x,\eta \right) }$,
\begin{equation}
n_{{\rm w}}=s_{{\rm w}}n=-\eta n,
\end{equation}
which is the topological quantum number. Now we see that both $\eta$ and $n$ are endowed
topological connotations, respectively representing the winding direction and the magnitude of winding number.

\subsection{Topological phase diagram}

\begin{figure}[t]
\includegraphics[width=1.0\columnwidth]{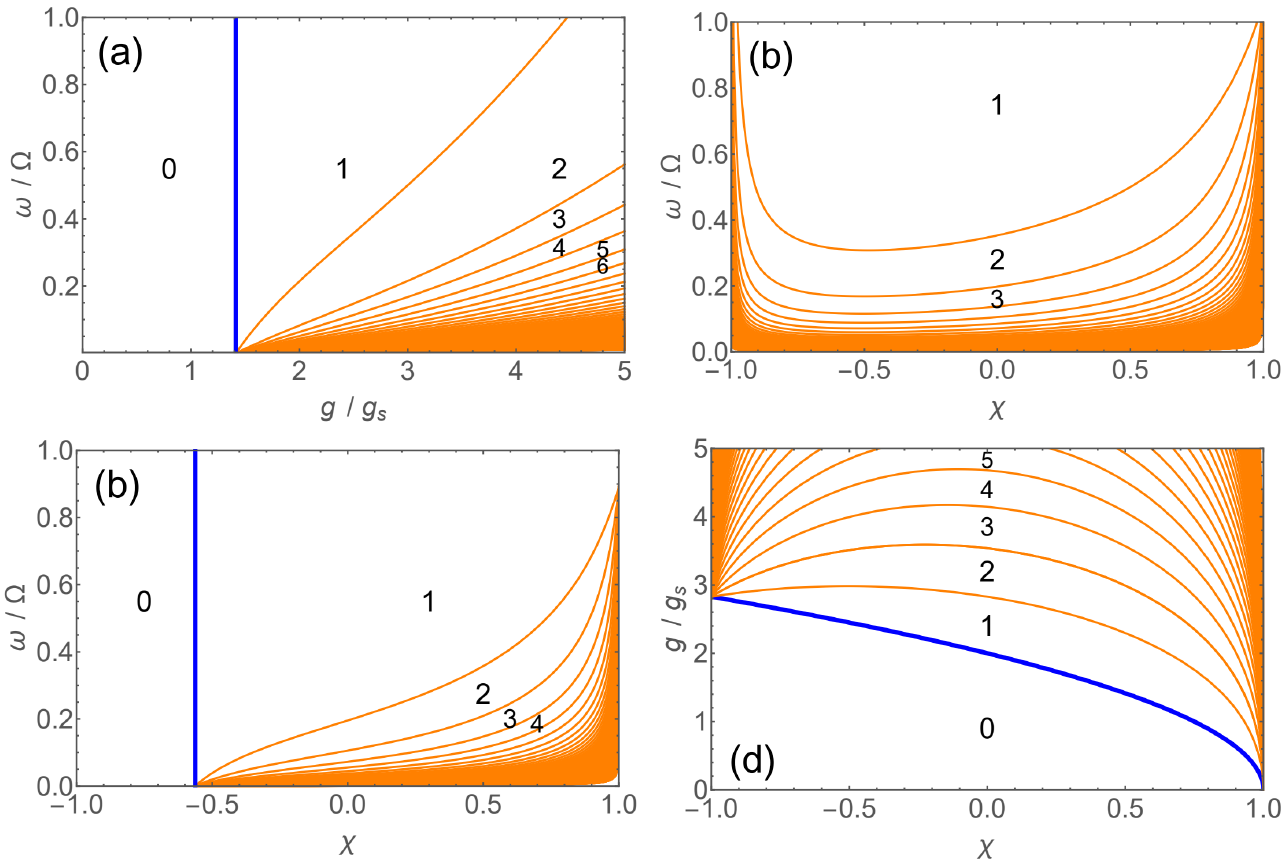}
\caption{{\it Topological phase diagrams. } Ground-state spin winding number
$n_{{\rm w}}$ in $g$-$\omega$ plane at $\chi =0.5$ (a), $\chi$-$\omega$ plane for $g=2.5g_{{\rm s}}$ (b) and
$g=3.0g_{{\rm s}}$ (b) at $\chi =0.5$, and $\chi$-$g$ plane at $\omega=0.3\Omega$ (d).}
\label{fig-phase-diagram}
\end{figure}

To have an overall view of all the phase transitions, we show the
ground-state phase diagrams in $g$-$\omega $, $\chi $-$\omega $, and $\chi $-$g$ planes by
Fig. \ref{fig-phase-diagram} where the numbers mark $n_{{\rm w}
}$. The phase boundaries shifting states from $\psi _{n}^{(x,-1)}$ to $\psi
_{n+1}^{(x,-1)}$ can be analytically obtained for $n=0$
\begin{equation}
g_{c}^{(0,1)}=2g_{\rm s}\sqrt{1-\chi },
\end{equation}%
and for $n>0$%
\begin{equation}
g_{c}^{(n,n+1)}=2g_{\rm s}\sqrt{\left( \chi _{+}+2n\chi _{+}\chi _{-}\right)
\widetilde{\omega }-\chi +S_{r}},
\end{equation}
where $S_{r}=\sqrt{(1-\chi _{+}\widetilde{\omega })^{2}+4n(n+1)\chi _{+}\chi
_{-}\widetilde{\omega }^{2}}$, $\widetilde{\omega }=\omega /\Omega $, $\chi _{\pm }=1\pm \chi $, and $g_{\rm s}=\sqrt{\omega \Omega}/2$.
Here in the figure the blue (thick) line
represents the principal boundary $g_{c}^{(0,1)}$ where the first transition
occurs from $n_{{\rm w}}=0$ phase to $n_{{\rm w}}=1$ phase when the coupling
$g$ is increasing at a fixed Stark coupling $\chi $ in panel (a) or when
$\chi $ is increasing at a fixed $g$ in panel (b). The principal boundary
disappears if the fixed $g$ is larger than $g_{c}^{(0,1)}$ as in panel (c),
which can be seen more clearly in panel (d) where $g_{c}^{(0,1)}$ exists in
a finite range within the physical regime $\chi \in \lbrack -1,1]$. Here one
finds that the critical couplings can be tuned by $\chi $.

\section{Simultaneous occurrence of Landau-class and topological-class
transitions}

\label{Sect-Landau-Topo}

\begin{figure}[t]
\includegraphics[width=1.0\columnwidth]{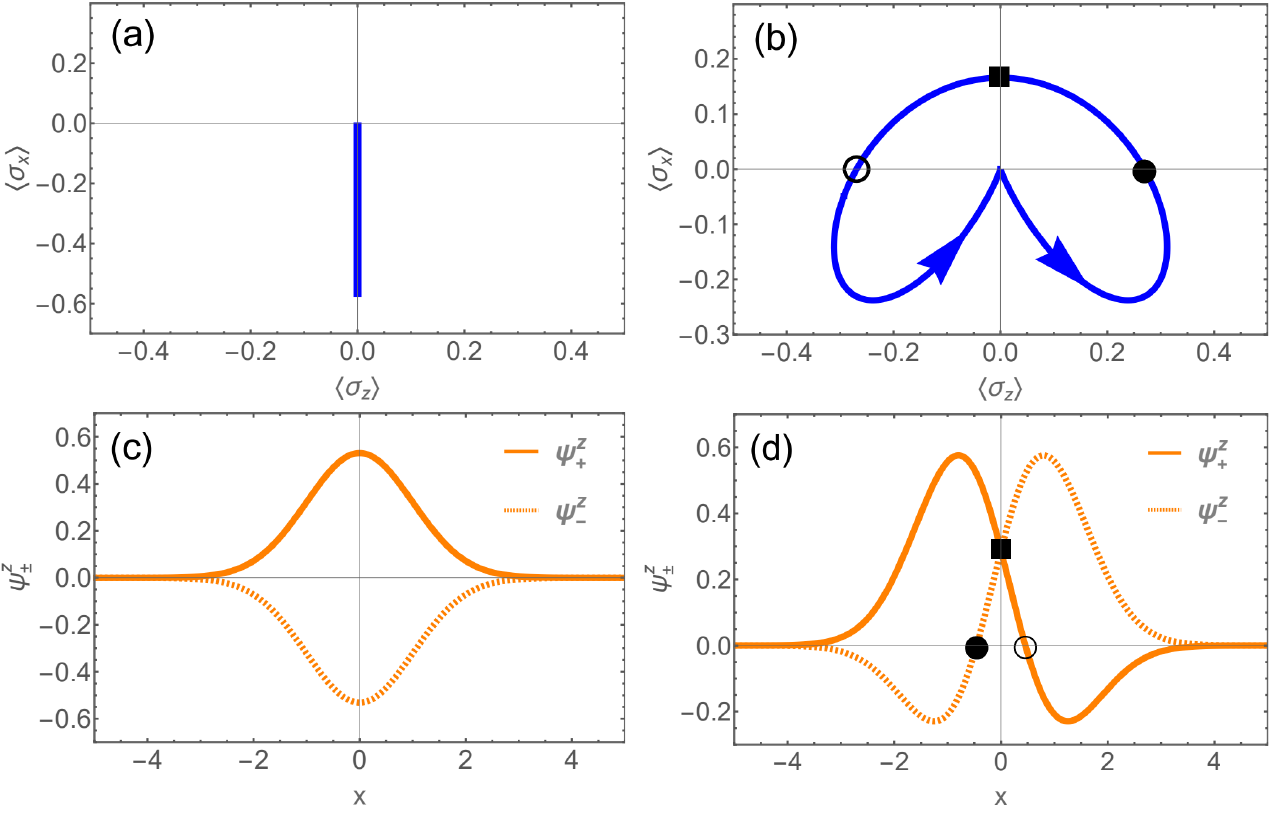}
\caption{{\it Simultaneous occurrence of symmetry-protected topological
class of transition and symmetry-breaking Landau class of transition.}
(a,b)
Ground-state spin profile in $\langle \sigma _{z}\left( x\right)
\rangle$-$\langle \sigma _{x}\left( x\right) \rangle$ plane in $n_{{\rm w}}=0$ phase (a) and $n_{{\rm w}}=1$ phase (b).
(c,d)
Ground-state wave function $\psi _{\pm}^z(x)$ with $\hat{P}_x$ and $\hat{P}_\sigma$ symmetries in $n_{{\rm w}}=0$ phase (c) and broken $\hat{P}_x$ and
$\hat{P}_\sigma$ symmetries in $n_{{\rm w}}=1$ phase (d). Here $g=0.8g_{{\rm s}}$ (a,c) and $g=2.0 g_{{\rm s}}$ (b,d) at $\omega =0.6
\Omega$, $\chi =0.5$. }
\label{fig-Landau-Topo}
\end{figure}

As mentioned in Introduction, the Landau class of transitions break the
symmetry while the topological class of transitions preserve the symmetry.
Conventionally these two classes of transitions are incompatible due to the
contrary symmetry requirements. However, here the principal transition at $%
g_{c}^{(0,1)}$ provides a paradigmatic case of exception, as it turns out to
be both Landau class of transition and topological class of transition
simultaneously.

\subsection{Topological-class transition feature at $g_{c}^{(0,1)}$:
Transition of spin winding topology}

As addressed in the previous section, we have seen the topological nature of
all the transitions, including the principal transition at $g_{c}^{(0,1)}$.
To have a direct feeling of the topological transtion for the principal
transition, in Fig. \ref{fig-Landau-Topo}(a,b) we plot spin profiles in $%
\langle \sigma _{z}\left( x\right) \rangle $-$\langle \sigma _{z}\left(
x\right) \rangle $ plane for the phases before and after the transition. As
one can see, in the $n_{{\rm w}}=0$ phase the spin profile is completely
flat ($\langle \sigma _{z}\left( x\right) \rangle=0$ $\forall x$) and does not wind at all as in panel (a), while in the $n_{{\rm w}}=1$
phase the spin is winding non-trivially as in panel (b). These two totally
different spin widing styles provide a sharp topological contrast for
recognization of the topological nature of transition.

\subsection{Landau-class transition feature at $g_{c}^{(0,1)}$: Symmetry
breaking of space inversion and spin reversion}

The $n_{{\rm w}}=0$ phase before the principal transition at $g_{c}^{(0,1)}$
is also special as it posseses more symmteries than the Hamiltonian. Indeed,
besides the parity symmetry, the state $\psi _{0}=\left\vert 0,\Downarrow
\right\rangle $ in this phase has symmetries of space inversion and spin
reversion:%
\begin{equation}
\hat{P}_{x}\psi _{0}=\psi _{0},\qquad \hat{P}_{\sigma }\psi _{0}=-\psi _{0},
\end{equation}%
where%
\begin{equation}
\hat{P}_{x}=\left( -1\right) ^{a^{\dagger }a},\qquad \hat{P}_{\sigma
}=\sigma _{x}.
\end{equation}%
In the position space on the $\sigma _{z}$ basis the wave function takes the
form $\psi _{0}=\psi _{0,+}^{z}\left( x\right) \left\vert \uparrow
\right\rangle +\psi _{0,-}^{z}\left( x\right) \left\vert \downarrow
\right\rangle =\left[ \phi _{0}\left( x\right) \left\vert \uparrow
\right\rangle -\phi _{0}\left( x\right) \left\vert \downarrow \right\rangle
\right] /\sqrt{2}$ where $\phi _{0}\left( x\right) $ is the Gaussian
function. The symmetry operator $\hat{P}_{x}$ actually inverses the space
\cite{Ying2020-nonlinear-bias} of a function $\hat{P}_{x}F\left( x\right)
=F\left( -x\right) $ which gives
\begin{eqnarray}
\hat{P}_{x}\psi _{0}\left( x\right) &=&\psi _{0}\left( x\right) ,
\label{Wave-0-Px} \\
\hat{P}_{\sigma }\psi _{0}\left( x\right) &=&-\psi _{0}\left( x\right) ,
\label{Wave-0-Pspin}
\end{eqnarray}%
as $\phi _{0}\left( x\right) $ is an even function. The space inversion and
spin reversion are more directly visible from the plot of the wave
function components in Fig. \ref{fig-Landau-Topo}(c). It should be mentioned
that theses symmetries in Eqs. \eqref{wave-Z-0}-\eqref{wave-Z-Inverse-x}
are rigorously fulfilled at any finite frequencies, in contrast to the QRM
and the anisotropic QRM where the low-frequency condition is required for
the validity of these symmetries~\cite{Ying-2021-AQT}. The unlimited frequency
condition greatly relaxes the experimental requirements for QPTs~\cite{Ying2015}.

On the contrary, in other phases with $n_{{\rm w}}\neq 0$, the symmetries of
space inversion and spin reversion are broken. Ineed, from Eqs. %
\eqref{wave-Z-0}-\eqref{wave-Z-Inverse-x} one can easily recognize%
\begin{eqnarray}
\hat{P}_{x}\psi _{n}^{\left( z,\eta \right) }\left( x\right) &\neq &\pm \psi
_{n}^{\left( z,\eta \right) }\left( x\right) , \\
\hat{P}_{\sigma }\psi _{n}^{\left( z,\eta \right) }\left( x\right) &\neq
&\pm \psi _{n}^{\left( z,\eta \right) }\left( x\right) ,
\end{eqnarray}%
in contrast to the symmetry-preserving Eqs. \eqref{Wave-0-Px} and
\eqref{Wave-0-Pspin}. In Fig. \ref{fig-Landau-Topo}(d) with $n_{{\rm w}}=1$
one sees directly that the wave function is asymmetric under either space
inversion or spin reversion.

Thus, the principal transition from state $\psi _{0}$ to state $\psi
_{1}^{\left( z,\eta \right) }$ is accompanied with the symmetry breaking of
both space inversion and spin reversion. This symmetry breaking feature
holds without approximation at any frequencies. In such a symmetry-breaking
sense, the the principal transition also belongs to the Landau class of
transition. Also, in the Landau theory the energy is expressed as a functional
of some order parameters. We leave the discussion in terms of variational
energy as a functional of the order parameters in symmetry breaking around
the transition in Appendix \ref{Appendix-Variationl-E}.

\subsection{Key for reconciliation of the two contrary transition classes:
Unbroken higher symmetry}

We have seen at $g_{c}^{(0,1)}$ the simultaneous occurrence of the
topological class of transition and the Landau class of transition which are
conventionally incompatible due to opposite symmetry requirements. The key
for their simultaneous occurrence or coexistence essentially lies in the
reconcilable situation that the symmetry which the topological class of
transition preserves is actually different from the symmetries which the
Landau class of transition breaks. Indeed, the symmetry that protects the
topological feature of the spin winding for the eigenstates in the
transitions is the parity symmetry $\hat{P}$, which comprises both the space
inversion and the spin reversal
\begin{equation}
\hat{P}=\hat{P}_{x}\hat{P}_{\sigma }.
\end{equation}%
As mentioned in Sect. \ref{Sect-Model}, the $g_{y}$ term in the coupling is effectively the
Rashba spin-orbit coupling or equal-weight mixture of the linear Dresselhaus
and Rashba spin-orbit couplings \cite%
{Ying-2021-AQT,Ying-gapped-top,Ying-Stark-top,Ying-Spin-Winding}, which
involves the spin nontrivially and drives the spin winding. The parity
symmetry guarantees the symmetric spin texture in \eqref{spin-symmetry} and its connection at the two infinity
ends in the position variation, which establishes the symmetry situation
for the TPTs. Note both before and after the transition $g_{c}^{(0,1)}$ this
parity symmetry that atcually protects all the TPTs is still preserved%
\begin{eqnarray}
\hat{P}\psi _{0}\left( x\right) &=&-\psi _{0}\left( x\right) , \\
\hat{P}\psi _{n}^{\left( z,\eta \right) }\left( x\right) &=&\pm \psi
_{n}^{\left( z,\eta \right) }\left( x\right) ,
\end{eqnarray}%
even when the subsymmetries in the space inversion and the spin reversal are
both broken. Therefore, the conventionally opposite symmetry requirements
for the Landau class and topological class of phase transitions reconcile
each other here and we see the simultaneous occurrence or coexistence of the
two contrary transition classes.

\section{Understanding unconventional topological transitions in the presence
of counter-rotating term}

\label{Sect-Unconv-TPTs}

\begin{figure*}[t]
\includegraphics[width=2.0\columnwidth]{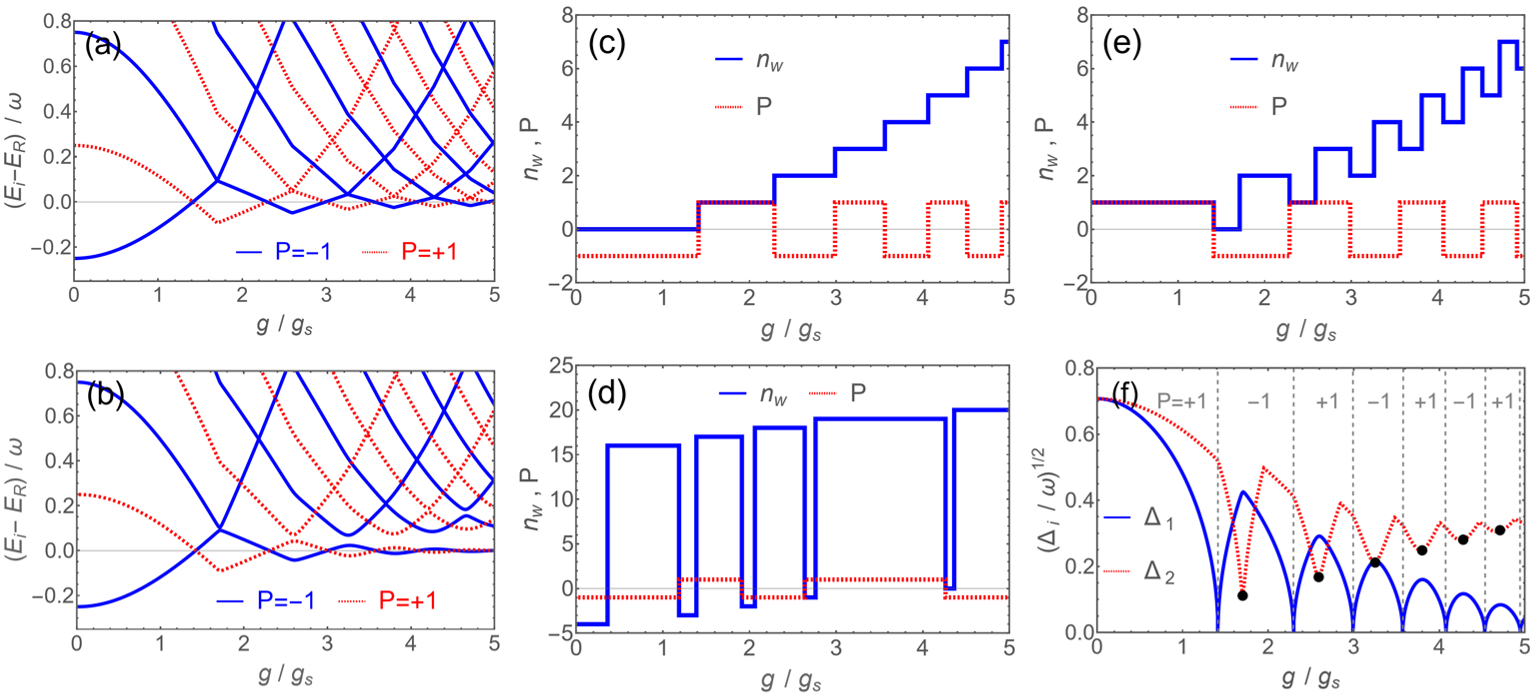}
\caption{
{\it Topological transitions without parity variation and origin for unconventional topological transitions without gap closing.}
(a,b)
Energy spectrum $E_i- E_R$ in the absence ((a), $\lambda =0$) and in the
presence ((b), $\lambda =0.01$) of the counter-rotating term, with the
reference energy $E_R=(E_2+E_1)/2$.
(c,d,e)
Spin winding number $n_{{\rm w}}$ (blue solid) and parity $P$ (red dotted) for $j_E=1$ (c), $j_E=20$ (d), and $j_E=2$ (e) at $\lambda =0$.
(f)
The first and second excitation gap $\Delta_i =E_{i+1}-E_{i}$ at $\lambda =0.01$. The dots in (f) are results by Eq. \eqref{perturbation-gap}. Here in all panels $\omega =0.3 \Omega$ and $\chi=0.5$.}
\label{fig-Unconv-TPTs}
\end{figure*}
\begin{figure*}[t]
\includegraphics[width=2.0\columnwidth]{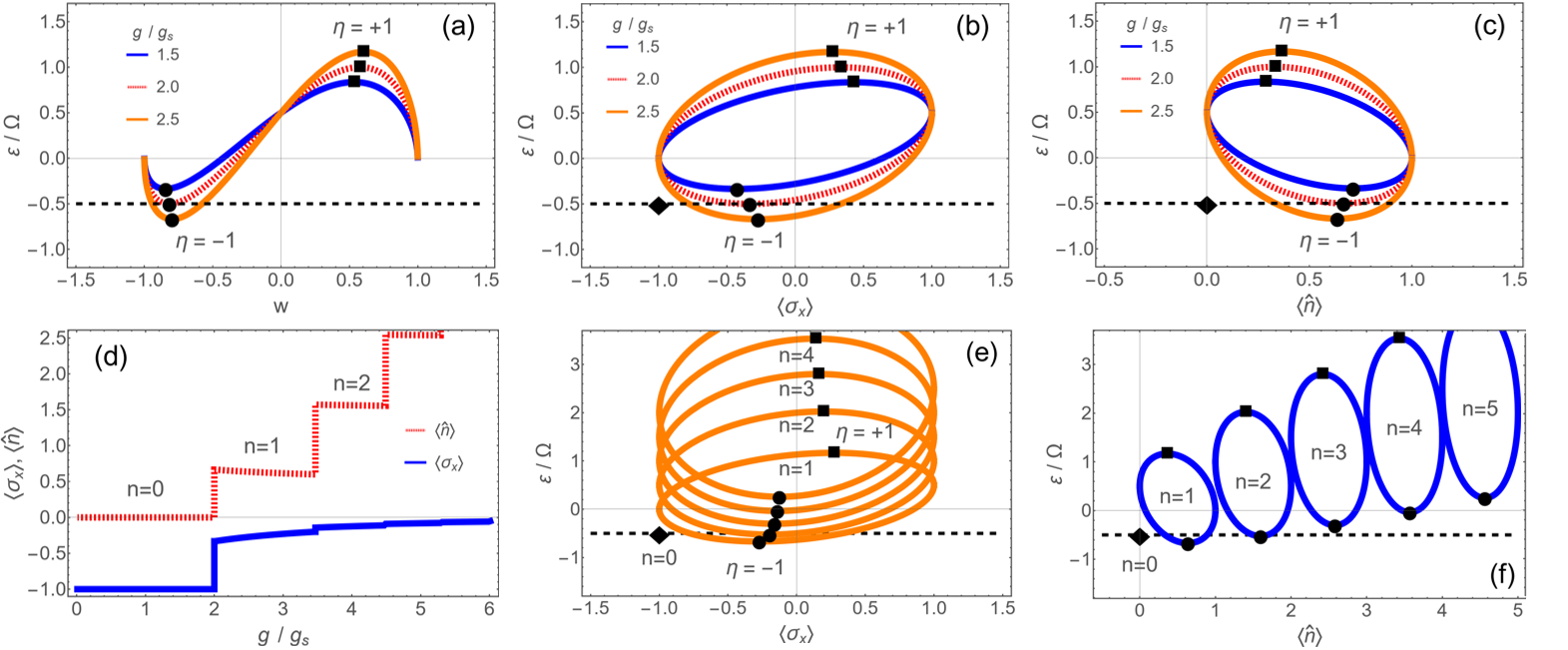}
\caption{
{\it Variational energy $\varepsilon$ as a functional of order parameters in symmetry breaking.}
(a-c)
$\varepsilon$ as a functional of $w$ (a), $\langle\sigma_x\rangle$ (b), $\langle\hat{n}\rangle$ (c) for state $\psi_1$ before ($g=1.5g_{{\rm s}}$, orange
(light) solid line), at ($g=2.0g_{{\rm s}}$, dotted line) and after ($g=2.5g_{{\rm s}}$, blue solid (dark) line) the transition $g_c^{(0,1)}$ in
competition with the energy of state $\psi_0$ (thin dashed line).
(d)
$\langle\sigma_x\rangle$ (solid) and $\langle\hat{n}\rangle$
(dotted) versus $g$ for the ground state. (e,f) $\varepsilon$ as a
function of $\langle\sigma_x\rangle$ (e) or $\langle\hat{n}\rangle$
(f) for states $\psi_n$ at $g=2.5g_{{\rm s}}$. The dots mark the
minimized energy and the squares label the maximized energy, which
reproduces the exact JC energy in $\eta =-1$ and $\eta =+1$
branches, while the diamond locates the order parameters for $\psi_0$. State $\psi_0$ (diamond) preserves the subsymmetries $\hat{P}_x$
and $\hat{P}_\sigma$ as indicated by integer numbers of $\langle \sigma_x\rangle$ and $\langle\hat{n}\rangle$, while the other states
(dots and squares) break the subsymmetries as reflected by deviations of $\langle\sigma_x\rangle$ from $\pm 1$ and $\langle\hat{n}\rangle$ from integer
numbers. Here $\omega=0.5\Omega$, $\lambda=0$, $\chi=0$ in all panels.
}
\label{fig-vari-E}
\end{figure*}

Most TPTs in the anisotropic QRM are conventional ones~\cite{Ying-2021-AQT}
which occur with gap closing as those in condensed matter~\cite%
{Topo-Wen,Hasan2010-RMP-topo,Yu2010ScienceHall,Chen2019GapClosing,TopCriterion,Top-Guan,TopNori}%
. Unconventional TPTs without gap closing also exist~\cite%
{Ying-gapped-top,Ying-Stark-top} analogously to the unconventional cases in
the quantum spin Hall effect with strong electron-electron interactions~\cite%
{Amaricci-2015-no-gap-closing} and the quantum anomalous Hall effect with
disorder~\cite{Xie-QAH-2021}. These unconventional TPTs lie in the ground
state by a mechanism of node introduction from the infinity~\cite{Ying-gapped-top,Ying-Stark-top}.
On the other hand, it is found that
unconventional TPTs emerge more frequently in excited states, especially
around level anti-crossings~\cite{Ying-Spin-Winding}. Here from the JC-Stark
model we can gain some insight for the origin of such unconventional TPTs in
excited states.

In Fig. \ref{fig-Unconv-TPTs} panel (a) shows the energy spectrum of the
JC-Stark model, where levels are crossing among the all states with negative
parity (blue solid) and positive parity (red dotted). Panel (b) gives the
spin winding number $n_{w}$ (solid line) of the ground state ($j_{E}=1$) which
jumps always at parity variation (dotted line) and gap closing (similar to
the solid line in panel (f)). The spin winding can turn direction in excited
states, as indicated by the negative values of $n_{w}$ in panel (d) for
state $j_{E}=20$ according to the discussion in Sect. \ref{Sec-nW-direction}. Each jump of $n_{w}$ is accompanied with a TPT.
In particular, some
TPTs occur without parity variation, as illustrated in panel (e)
for state $j_{E}=2,$ which comes from the level crossing between same parity
states. Note here the gap is sill closing at the transitions despite no
parity variation. What interesting is, once we add anisotropy, e.g. $\lambda
=0.01$, the gap at these TPTs opens, as demonstrated by the energy
spectrum by panel (b) and more clearly by the dotted line in panel (f). This
gap opening accounts for the afore-mentioned unconventional TPTs in the
excited.

A clearer explanation can be given by regarding the counter-rotating term as
a perturbation around these transitions which mainly involves the two level-crossing states $\psi
_{n}^{\left( x,\eta \right) },\psi _{n^{\prime }}^{\left( x,\eta^{\prime } \right) }$
with energies $\varepsilon _{1}=E^{\left( n,\eta \right) }$, $\varepsilon
_{2}=E^{\left( n^{\prime },\eta^{\prime } \right) }$ and winding numbers $n_{w,1}=-\eta n$, $n_{w,2}=-\eta ^{\prime
}n^{\prime }$. On the basis of these two states the Hamiltonian in matrix
form can be written as
\begin{equation}
H\approx\left(
\begin{array}{cc}
E^{\left( n,\eta \right) } & d \\
d & E^{\left( n^{\prime },\eta ^{\prime }\right) }%
\end{array}%
\right)
\end{equation}%
where
\begin{eqnarray}
d =\langle \psi _{n}|H_{\lambda }|\psi _{n^{\prime }}\rangle
&=&
\lambda g
\sqrt{n^{\prime }+1}\frac{C_{n\Uparrow }^{\left( \eta \right) }C_{n^{\prime
}\Downarrow }^{\left( \eta ^{\prime }\right) }}{\sqrt{N_{n}N_{n^{\prime }}}}%
\delta _{n,n^{\prime }+2}  \nonumber \\
&&+\lambda g\sqrt{n^{\prime }-1}\frac{C_{n\Downarrow }^{\left( \eta \right)
}C_{n^{\prime }\Uparrow }^{\left( \eta ^{\prime }\right) }}{\sqrt{%
N_{n}N_{n^{\prime }}}}\delta _{n,n^{\prime }-2} \label{eq-d}
\end{eqnarray}%
and $H_{\lambda }$ is the counter-rotating term in \eqref{H-Lambda} beyond the
JC-Stark model. The crossing levels are split as $E_{\pm }=\frac{1}{2}%
(E^{\left( n,\eta \right) }+E^{\left( n^{\prime },\eta ^{\prime }\right)
}\pm \Delta )$ with a gap opening at the level-crossing point $E^{\left(
n,\eta \right) }=E^{\left( n^{\prime },\eta ^{\prime }\right) },$
\begin{equation}
\Delta =\sqrt{(E^{\left( n,\eta \right) }-E^{\left( n^{\prime },\eta
^{\prime }\right) })^{2}+4d^{2}}\rightarrow 2|d|, \label{perturbation-gap}
\end{equation}%
which is finite for $n=n^{\prime }\pm 2$, leading to the level
anti-crossing. The validity of Eq. \eqref{perturbation-gap} is confirmed by the dots which match well
the numerical result by exact diagonalization~\cite{Ying2020-nonlinear-bias,Ying-Spin-Winding} in the dotted line in Fig. \ref{fig-Unconv-TPTs}(f).
Here from Eqs. \eqref{eq-d} and \eqref{perturbation-gap} one sees that the gap opening does not occur for
crossing states with different parity, since they respectively have even and
odd $n$ as indicated by $P=\left( -1\right) ^{n}$ from Eq. %
\eqref{Eigen-Parity}. Note the small $\lambda $ here is a perturbation which
is not yet enough to change the winding numbers so that $n_{w}$ remains
similar to $\lambda =0$ case in Fig. \ref{fig-Unconv-TPTs}(e). Thus, the
TPTs originally at level crossing now become unconventional TPTs without gap closing as
the gap is opening. Larger $\lambda $ may induce more unconventional TPTs
than those inherited from the $\lambda =0$ case at the gap opening \cite%
{Ying-Spin-Winding}.  Finally it should be noticed that such unconventional
TPTs are still protected by the parity symmetry as the added term $%
H_{\lambda }$ preserves the parity symmetry, $[H_{\lambda },\hat{P}]=0$.
The above analysis provides a simple but clear understanding for the unconventional TPTs in excited states.

\section{Conclusions and Discussions}

\label{Sect-Conclusions}

We have presented a rigorous study to show the topological nature of
transitions in Jaynes-Cummings Model generally with Stark non-linear
Coupling, which is a fundamental model for light-matter interactions. The
exact and explicit solution of the model enables us to analytically analyze
the nodes of eigen wave functions and establish the exact correspondence to
the nodes in the spin texture. In the light of the Hermite polynomial
properties, we have proven that the spin nodes on $\langle \sigma _{z}\left(
x\right) \rangle $ and $\langle \sigma _{x}\left( x\right) \rangle $ axes
are interlacing on positive and negative axes, thus the node sequence forms
a smooth spin winding without anti-winding nodes. In particular, the
spurious fractional winding angle at infinity is found to be integer, which
achieves a full winding. Thus, the phase transitions in the model have a
nature of TPTs.

Based on a strict derivation we have reformulated the spin winding number to facilitate
the extraction of winding numbers by replacing the integral formula with an
algebraic formula in terms of finite points of nodes, which also bridges the geometrical topology
and the algebraic topology in a physical way. The excitation number
and the energy branch label of eigen states turn out to the magnitude and the sign (winding direction)
of the winding number, thus both endowed a topological connotation unrecognized before.

In
particular, we have found that the $\langle \sigma _{x}\left( x\right) \rangle $
nodes are invariant, which might have potential advantage in designing
topological devices as they provide robust topological information
unaffected by variations of the parameters.

We have also demonstrated that the principal transition has the character of
Landau class of phase transition besides that of TPT, by pointing out the
symmetry breaking aspect and variational energy analysis as functional of
order parameters. Note conventionally Landau class of phase transitions and
topological class of phase transitions are incompatible due to the contrary
symmetry requirements, here the principal transition establishes a
paradigmatic case that a transition can be both symmetry-breaking Landau
class of transition and symmetry-protected topological class of transition
simultaneously. The key for the reconciliation of the two contradictory
classes of transitions lies in the preserved higher symmetry which protects the TPTs despite the
subsymmetries are broken in the Landau class of transition.

Moreover, we have applied our result to analyze the gap opening at some particular TPTs without parity variations in the
presence of the counter-rotating term, which gives an analytical explanation
for the unconventional TPTs without gap closing. Note that a gapped situation can avoid the detrimental time divergent problem in preparing sensing state~\cite{Ying2022-Metrology}, the unconventional TPTs may similarly have potential advantages in possible applications or designing quantum topological devices. In such a favorable situation, our understanding
might be helpful for further exploring and exploiting unconventional TPTs in light-matter interactions.

Finally it is worthwhile to mention that the model considered in the present
work may be implemented in realistic systems, e.g. in superconducting
circuits. Indeed, both the anisotropy~\cite{PRX-Xie-Anistropy,Forn-Diaz2010,Pietikainen2017,Yimin2018,Wang2019Anisotropy}
and the Stark nonlinear coupling~\cite{Stark-Grimsmo2013,Stark-Grimsmo2014,Stark-Cong2020} are adjustable.
Besides realizations of ultra-strong couplings in $\lambda \neq 0$ case \cite%
{Diaz2019RevModPhy,Wallraff2004,Gunter2009,Niemczyk2010,Peropadre2010,FornDiaz2017,Forn-Diaz2010,Scalari2012,Xiang2013,Yoshihara2017NatPhys,Kockum2017,Bayer2017DeepStrong}%
, access to ultra-strong couplings can be also possible for $\lambda =0$
\cite{Ulstrong-JC-1,Ulstrong-JC-3-Adam-2019,Ulstrong-JC-2}. The position $x$ can be represented by
the flux of Josephson junctions and the spin texture might be measured by
interference devices and magnetometer \cite{you024532}. These systems may
provide platforms for possible tests or applications of our results. Our
analysis might be also relevant for some other systems as the effective
Rashba/Dresselhaus spin-orbit coupling in our model shares similarity with
those in nanowires~\cite%
{Nagasawa2013Rings,Ying2016Ellipse,Ying2017EllipseSC,Ying2020PRR,Gentile2022NatElec},
cold atoms~\cite{Li2012PRL,LinRashbaBECExp2011} and relativistic systems~%
\cite{Bermudez2007}.

\section*{Acknowledgements}

This work was supported by the National Natural Science Foundation of China
(Grants No.~11974151 and No.~12247101).

\appendix

\section{Variational energy as functional of order parameters}
\label{Appendix-Variationl-E}

In this appendix we present some discussions in terms of energy functional
of order parameters in the situation of the symmetry breaking at the
principal transition. Although the exact solution has been obtained in Sec. %
\ref{Sect-Solution}, a reformulation for energy as a functional of order
parameters is more connected with the Landau theory of phase
transitions.\ Under the constraint of the $U(1)$ symmetry the eigenstate of
the JC-Stark model should be either a linear combination of
bases $\left\vert n-1,\Uparrow \right\rangle $ and $\left\vert n,\Downarrow
\right\rangle $
\begin{equation}
\psi _{n}=\sqrt{1-w^{2}}\left\vert n-1,\Uparrow \right\rangle +w\left\vert
n,\Downarrow \right\rangle
\end{equation}%
or composed solely of $\psi _{0}=\left\vert 0,\Downarrow \right\rangle .$
The energy of $\psi _{0}$ is simply $E^{0}=-\Omega /2$, while the energy of $%
\psi _{n}$ is variational with respect to the basis weight $w$:
\begin{equation}
\varepsilon =\varepsilon _{0}+\left[ \omega -\Omega -\chi \omega \left(
2n-1\right) \right] w^{2}+2\sqrt{n}g\sqrt{1-w^{2}}w.
\end{equation}%
where $\varepsilon _{0}=\left( 1+\chi \right) \omega \left( n-1\right) +%
\frac{\Omega }{2}$ is independent of $w$. The minimization and maximization
of $\varepsilon $ with respect to $w$ lead to%
\begin{equation}
w_{\pm }=\pm \sqrt{\frac{4g^{2}n+A_{w}^{2}\pm A_{w}\sqrt{4g^{2}n+A_{w}^{2}}}{%
8g^{2}n+2A_{w}^{2}}},
\end{equation}%
where $A_{w}=[(2n-1)\chi -1]\omega +\Omega ,$ which is equivalent~\cite{YingVarM} to %
\eqref{Cx-up}\eqref{Cx-down} with $\eta =\pm $.

Note the relations
\begin{eqnarray}
\langle \hat{n}\rangle &=&n-1+w^{2}, \\
\langle \sigma _{x}\rangle &=&\left( 1-2w^{2}\right) ,
\end{eqnarray}%
the variational energy can be rewritten into a functional of the order
parameter $\langle \sigma _{x}\rangle $ or $\langle \hat{n}\rangle $%
\begin{eqnarray}
\varepsilon &=&\varepsilon _{0}-C_{\varepsilon }\left( 1-\langle \sigma
_{x}\rangle \right) +\eta \sqrt{n}g\sqrt{1-\langle \sigma _{x}\rangle ^{2}}
\\
\varepsilon &=&\varepsilon _{0}-2C_{\varepsilon }\left( \langle \hat{n}%
\rangle +1-n\right)  \nonumber \\
&&+\eta 2\sqrt{n}g\sqrt{\left( n-\langle \hat{n}\rangle \right) \left(
\langle \hat{n}\rangle +1-n\right) }
\end{eqnarray}%
where $C_{\varepsilon }=\left[ \frac{\Omega -\omega }{2}+\left( n-\frac{1}{2}%
\right) \chi \omega \right]$. Fig. \ref{fig-vari-E} illustrates some
examples of the variational energy around the transition at $g_{c}^{(0,1)}$
in competition to the energy of $\psi _{0}$. The dots mark the minimized
energy while the squares label the maximized energy. One sees that when $g$
is increasing, the minimized energy becomes lower than that of state $\psi
_{0}$, which triggers a first-order transition unlike the second-order
transition in the QRM or the anisotropic QRM \cite%
{Ying2020-nonlinear-bias,Ying-Stark-top} in the low frequency limit. The
upper branch ($\eta =+1$) and the lower branch ($\eta =-1$) of the
variational energy $\varepsilon $ are connected and form energy circles as
shown in panel (d), the lowest and highest points are the final energies,
which together with $E^{0}$ reproduce the exact energies in Eq. (\ref{E-n-JC}%
) not only for the ground state but also for the excited states as
demonstrated in panels (e) and (f).

Note the energy and the order parameters of $\psi _{0}$ are represented by
the diamonds in Fig. \ref{fig-vari-E}(b,c,d,e), where the expectation value $%
\langle \sigma _{x}\rangle =\langle \hat{P}_{x}\rangle =-1$ implies the spin
reversal symmetry in (\ref{Wave-0-Px}) and $\langle \hat{n}\rangle =0$
indicates the space inversion symmetry in (\ref{Wave-0-Pspin}). As a
contrast, the values of $\langle \sigma _{x}\rangle $ and $\langle \hat{n%
}\rangle $ for the minimum and maximum points on the energy circles deviate
from the integer numbers, which means breaking of these symmetries.

\end{document}